\documentclass{article}
\usepackage[utf8]{inputenc}
\usepackage{cite}
\usepackage{graphicx} 
\usepackage{tcolorbox} 
\usepackage{csquotes}
\usepackage{hyperref}
\usepackage{rotating}
\usepackage{pdfpages}

\usepackage{amsthm}
\theoremstyle{plain}
\newtheorem{risk}{Identified Risk}
\newtheorem*{risk*}{Identified Risk 0}

\usepackage[font=footnotesize,labelfont=bf]{caption}

\usepackage{todonotes}

\title{Systemic Risks of Interacting AI}
\author{Paul Darius$^1$, Thomas Hoppe$^1$, Andrei Aleksandrov$^1$}
\date{$^1$Fraunhofer FOKUS, Berlin, Germany}

\begin{document}
\includepdf{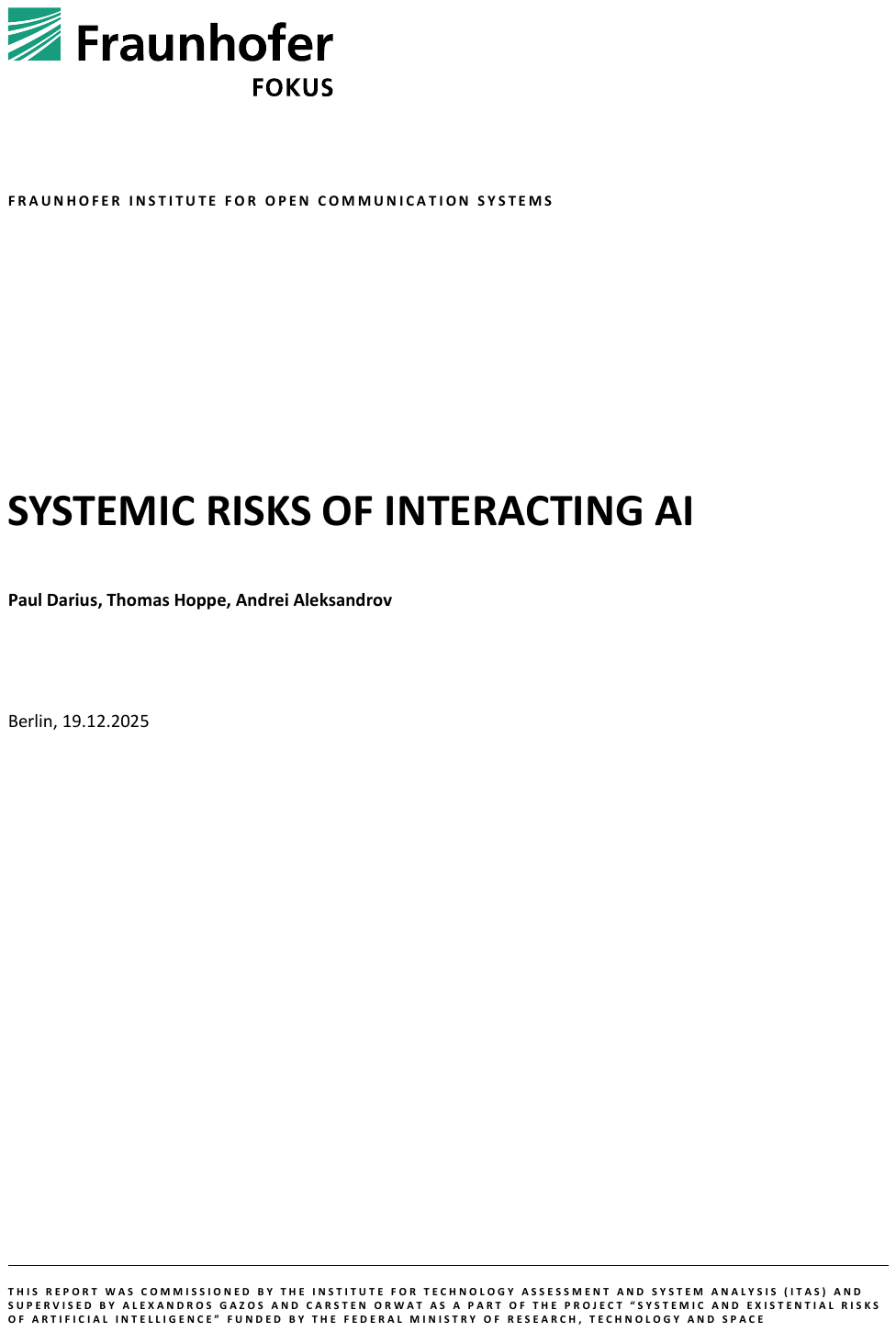}

\maketitle

\begin{abstract}
In this study, we investigate system-level emergent risks of interacting AI agents. The core contribution of this work is an exploratory scenario-based identification of these risks as well as their categorization. We consider a multitude of systemic risk examples from existing literature and develop two scenarios demonstrating emergent risk patterns in domains of smart grid and social welfare. We provide a taxonomy of identified risks that categorizes them in different groups. In addition, we make two other important contributions: first, we identify what emergent behavior types produce systemic risks, and second, we develop a graphical language \enquote{Agentology} for visualization of interacting AI systems. Our study opens a new research direction for system-level risks of interacting AI, and is the first to closely investigate them.
\end{abstract}

\tableofcontents
\newpage

\section{Introduction}

The development of artificial intelligence (AI) since the beginning of this millennium has led to the creation not only of intelligent computer vision, machine translation and speech understanding tools, but also of generative systems producing text, images and videos on demand. Recent developments indicate that agentic AI, or more broadly \emph{interacting AI}, consisting of AI components interacting with each other may solve harder problems and relieve humans from mundane tasks. Unfortunately, the misuse of agentic AI can similarly lead to significant damage and harm. If potential consequences of a misbehaving system can affect a large part of a society by harming social values, assets or goods, we are speaking of a \emph{systemic risk}.

Systemic risks come in different varieties and their potential harm can be spread through society in different ways. Some systemic risks can cause sudden catastrophic consequences. For example, a mistake in a decentralized AI-driven smart grid can cause a widespread energy blackout leading to complete halt of a functioning society. There are also systemic risks that capture slow but large-scale negative erosion processes within society. A prominent example is the domain of social media where the complex relationship between media, recommendation and ranking algorithms, and opinion dynamics leads to consistently increasing social polarization~\cite{santos2021link}. 

In this study, we focus on systemic risks that are \emph{emergent} in nature, the risks that arise from full complexity of interacting AI involving multiple \emph{AI agents} as well as communication between them. One can easily imagine that AI agents acting autonomously may be a source of unwanted, not anticipated effects emerging because of their interaction with each other. Hence, the questions of how exactly such emergent behaviour can arise in interconnected systems of intelligent agents and whether emergent behaviours have effects on large segments of society need to be addressed to identify mitigating measures in advance. 

Currently, we are standing at the beginning of research into systemic risks arising from interacting AI. This work uses scenario-based exploratory approach to identify these risks in different domains as well as proposes a taxonomy of them. For this purpose, we developed a graphical framework \enquote{Agentology} for illustration of emergent patterns within interacting AI systems. In addition, we investigated what emergent behavior types from the classification of Fromm~\cite{Fromm05} are associated with risks of interacting AI.

\subsection{Scope of the Study}

This study addresses the question of systemic risks that may emerge from interacting agents as well as mechanisms of their emergence. Depending on the domain, the term \enquote{risk} has several meanings but is always connoted with a negative effect of an event~\cite{xu2024multidisciplinary}. Originating from the field of finance, the term \enquote{systemic risk} is used to denote the possible collapse of an entire financial system or market~\cite{schweizer2019governance,renn2022systemic}. Hence, the term \enquote{systemic} implies an event with a large extend of consequences, even so its occurrence might not be very probable.

Risks depend on events, and events are domain-dependent, implying that risks are domain dependent too. For example, a collapse of the entire financial system will have a certain interaction structure, which probably will not occur in any other domain. As another example, let us consider an emerging pandemic that is caused by high-infectious viruses distributed by global travels without available vaccines. A similar structure may occur in computer networks for the distribution of new yet unknown computer viruses. Simultaneously, pandemics and computer viruses can be the only two domains with this specific risk structure.

In the domain of information technology, \enquote{agents} denote active software components which fulfill certain functions~\cite{balaji2010introduction}. Such functions may range from simple tasks like controlling the temperature of a heating system, over searching the World Wide Web for information, navigating humans to a goal, up to driving autonomously a car. Such software agents can be \emph{deterministic}, like a heating controlling agent, or \emph{probabilistic}, in the sense that the outcome of an action cannot be predicted in advance, because of the huge amount of internal parameters, external influences, potential states and the unknown current state of the agent~\cite{russell2016intelligent}. However, even humans or organisations are subsumed under the term agent~\cite{daedalus}, although they are probabilistic in the sense above, they additionally may be acting rational or irrational.

Any systemic risk is always dependent on a particular domain or overlapping domains. Hence, any classification of systemic risks of interacting agents needs to consider only negative events, and address detrimental or fatal events, affecting the entire system or larger subsystems thereof. In addition, such classification should cover events, which emerge from the interaction of more than one active agent, and consider interactions between different forms of deterministic, probabilistic, rational and irrational agents. This approach will be followed throughout this work.

\subsection{Related Studies}

Here, we want to survey studies that pursue goals similar to ours and compare our contributions to these studies. This work is defined by three central concepts: \emph{artificial intelligence}, \emph{systemic risk}, and \emph{interaction between AI agents}. To our knowledge, existing studies do not cover all three together.
For starters, there are several works that focus on systemic risks without closely considering AI~\cite{schweizer2019disaster,ortwin2022systemic,oecd2025managing} as well as a number of studies focused on systemic AI risks without considering interacting agents~\cite{hendrycks2023overviewcatastrophicairisks,turchin2020classification}.
We consider the work by Hammond et. al.~\cite{hammond2025multiagentrisksadvancedai} as the closest to our study. It explores many different risks of multi-agent AI systems and aspects that underlie these risks, and develops a simple categorization for them. While there is a significant overlap between risks considered in this study and in Hammond et. al.~\cite{hammond2025multiagentrisksadvancedai}, we generalize many of the risks they present, consider new risks such as agent exit risks, and have strong focus on systemic consequences. 
Another study by Manheim~\cite{manheim2019multiparty} considers abstract risk scenarios of multi-agent systems but does not evaluate their systemic impact. 
A study by Mogul~\cite{mogul2006emergent} considers risks of systems with interacting agents, but does not cover full scope of artificial intelligence and does not consider systemic impact.
An article by Raza et. al.~\cite{raza2025trismagenticaireview} considers technical risks of multi-agent LLM-based systems. The presented risks can be a part of a cause for a systemic risk, but this is not considered in their article.
Also, there are many studies that present concrete experiments to demonstrate risk scenarios but compared to our study, do not provide an overview across multiple domains or a taxonomy. Many of such studies are covered in our risk identification literature review in Section~\ref{sec:Literature_Review}.
To our knowledge, this study is the first that focuses on the full combination of all three mentioned aspects: artificial intelligence, systemic risk and AI agents' interaction.

\subsection{Contributions and Structure of the Study}

The main goal of the study is identification of emergent systemic risks across different domains in contexts involving interacting AI agents. To highlight identified risks in the text and enable a convenient cross-referencing in the paper, each risk will be presented in the following form:
\begin{risk*}[Risk Name]
An abstract description of the identified risk.
\end{risk*}
Our study is structured in multiple sections. In Section~\ref{sec:Background}, we establish basic concepts such as artificial intelligence and multi-agent systems. We develop basic terminology used throughout the article in Section~\ref{sec:Definitions}. In the sections afterwards, we present main contributions of our paper as listed below.
\begin{itemize}
    \item \textbf{Graphical notation for interacting AI.} In Section~\ref{sec:Agentology}, we present a structured graphical notation \enquote{Agentology} for modeling interacting AI systems. Agentology includes two types of diagrams helping illustrate behavior patterns within a system.
    \item \textbf{Risk in Fromm's taxonomy of emergent behavior.} Fromm's typology~\cite{Fromm05} is a standard taxonomy of emergent behavior. We identify which of the types are relevant to systemic risks of interacting AI and supply a couple of risk scenarios in Section~\ref{sec:MAS_Structures}.
    \item \textbf{Systemic risk patterns in existing literature.} In Section~\ref{sec:Literature_Review}, we survey existing literature for emergent systemic risk scenarios.
    \item \textbf{Scenario-driven risk identification.} Section~\ref{sec:Case_Studies} contains two scenarios with emergent risk patterns in domains of smart grid and social welfare. The scenarios are constructed based on recent developments in the science of intelligent systems, and help us identify several systemic risks.
    \item \textbf{Taxonomy of emergent systemic risks.} In Section~\ref{sec:Taxonomy_of_Systemic_Risks}, we present the taxonomy categorizing risks identified within previous sections.
\end{itemize}
Our study concludes with a summary and future work proposals in Section~\ref{sec:Conclusion}.

\section{Background}
\label{sec:Background}

\subsection{Artificial Intelligence}
The initial goal of artificial intelligence (AI) was  originally formulated as to understand whether “every aspect of learning or any other feature of intelligence can in principle be so precisely described that a machine can be made to simulate it”\footnote{McCarthy, J., Minsky, M., Rochester, N., Shannon, C.E., "A Proposal for the Dartmouth Summer Research Project on Artificial Intelligence"., http://raysolomonoff.com/dartmouth/boxa/dart564props.pdf, August 1955} After an initial phase of disillusionment in the 1960s, AI experienced its first upswing from the end of the 1970s. Knowledge-based approaches based on symbolic logic and promoted in the USA, Europe and Japan led to the first solutions.

After a second AI winter, the beginning of the digital age - the point at which the amount of digital data surpassed that of analog data \cite{HilberLiopez2011} - led to a revival of AI research~\cite{RussellNorvig2010}. Thanks to the internet, large amounts of data and server capacities were available for intensifying work on neural networks, particularly for machine learning, so that the foundations could be laid for breakthroughs in image and speech processing, text analysis for search engines, and playing chess and Go in the 2010s. The development of transformers, large language models and generative image and video generation in the early 2020s brought AI to the attention of broad sections of the media, companies and the general public. Nowadays, AI is therefore almost exclusively equated with machine learning.

Knowledge-based AI was driven forward with little public attention, particularly by the concepts of the “Semantic Web” and “Linked Data” proposed by Tim Berners-Lee. The aim was to make the content of the internet easier for machines to process, but the actual benefit of these approaches was more evident in corporate use when integrating separate data silos.

\begin{figure}[t]
  \centering
  \includegraphics[width=1\linewidth]{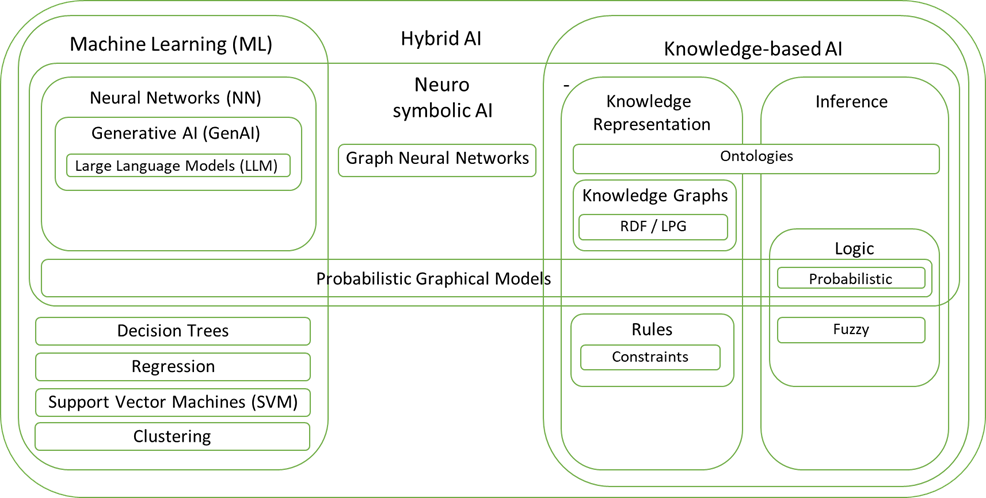}
  \caption{Landscape of AI Approaches}
  \label{fig:Landscape of AI}
\end{figure}

AI approaches can therefore essentially be divided into three categories. Knowledge-based AI, machine learning, which also includes all forms of neural networks, and hybrid approaches \cite{HinkelmannEtAl2025}. The Figure~\ref{fig:Landscape of AI} illustrates these relationships and overlaps.

\subsubsection{Knowledge-based AI}

The foundations of knowledge-based AI approaches are based on the symbolic representation of models and their mostly logic- or graph-based processing. What is essential about them is that the representing symbols are usually of a digital, i.e. clearly delineated, nature. By means of formally clearly defined semantics of the representation language used, logical inferences and graph algorithms, the meaning of the symbols, the conclusions drawn from them and the derived information can generally be explained.

These approaches follow the principle of “knowledge-basedness” in which the represented models are distinguished from the actual processing mechanisms. This allows the latter to be reused for other areas of knowledge. The models are usually developed by humans or automatically derived from other information. If no heuristic processing methods are used, these procedures are generally deterministic.

\subsubsection{Machine Learning}

Machine learning methods are largely based on numerical, sub-symbolic representations of information and are characterized by a separation between two processing phases: the training of models and their subsequent use. Training is usually based on a large amount of data that is as representative as possible. The model derived from this generalizes this data so that it can also be applied to previously unknown data of the same type in the usage phase.

The numerical representation is referred to as sub-symbolic, as the meaning, particularly of continuous numerical values, is not clearly defined and can also be distributed across several properties of the data.

Due to the training data in particular, these methods are often subject to the problem of bias, i.e. the conscious or unconscious favoring or disfavoring of certain values when selecting the training data. 

From the perspective of using the models, machine learning methods are often non-deterministic, as they are subject to some probabilistic effects. The selection of training data is to a certain extent subject to chance. The methods themselves often sort the data randomly during training, so that the resulting models and the information derived from them also exhibit a certain degree of indeterminacy during use. If the models are further trained using new data or if the applications implemented with them are self-learning, the derivation of the same predictions for older data cannot be guaranteed. 

\textbf{Neural Networks} represent a subset of machine learning that attempts to translate the functional principles of the human nerve cell into a machine processing method. In this process, a large number of entities (referred to as “neurons”) organized in layers are modified by the values of an input layer until they produce the values of an output layer. The individual neurons consist of a numerical value and the neurons of one layer are connected by weightings to all neurons of the respective subsequent layers. Input values lead to a change in the activation of the subsequent layers through so-called activation functions and deviations of the output from the expected output through backward propagation of the errors to adjust the weightings.

Although this processing paradigm is generally strictly deterministic, the learned models are randomly dependent due to different processing sequences and accuracy-related limitations of numerical values.

\textbf{Deep Learning} summarizes different architectures based on the above processing  principle of neural networks, which differ in the number of layers and neurons in the layers, activation functions, data type of the weightings, feedback from layers and memory retention.  In particular, the class of recurrent neural networks and transformers have created the prerequisites for the class of generative AI methods, to which large language models also belong.

\textbf{Generative AI} is the subsuming term for all deep learning systems generating text, images videos or in the future also 3D-scenes. With the release of ChatGPT, \textbf{Large Language Models (LLM)} became well known. Although LLMs,   general purpose AI models and agent-based AI are the current hot topics, they still have weaknesses when it comes to understanding the world, using more complex logical reasoning or mathematical understanding. Leading scientists, like Yann LeCun\footnote{(Lu, 2018) “Yann LeCun’s IJCAI Keynote: We Need a World Model“, Victor Lu, Juli 2018, syncedreview in medium.com https://medium.com/syncedreview/yann-lecuns-ijcai-keynote-weneeda-world-model-b5248f50e50f}, Doug Lenat \& Gary Marcus\footnote{(Lenat \& Marcus, 2023) “Getting from Generative AI to Trustworthy AI: What LLMs might learn from Cyc”, Doug Lenat, Gary Marcus, arXiv:2308.04445, Juli, 2023} or Stephen Wolfram\footnote{(Wolfram, 2023) “Wolfram Alpha as the Way to Bring Computational Knowledge Superpowers to ChatGPT”, Steven Wolfram, January 2023, in: Steven Wolfram Writings, https://writings.stephenwolfram.com/2023/01/wolframalpha-as-the-way-to-bring-computational-knowledgesuperpowersto-chatgpt}, in the fields of machine learning, neural networks, and knowledge representation see the solution to these difficulties in the combination of trained models with curated world knowledge formalized by knowledge-based methods. Such combinations called neuro-symbolic or hybrid AI are already under development and in practical use.

\subsection{Multi-Agent System}
A \textbf{Multi-Agent System (MAS)} is a computer-based system composed of several interacting intelligent agents. Such intelligent agents are supposed to perceive their environments, act autonomously for a certain period to achieve goals and may improve their performance by learning or through the acquisition of knowledge. Autonomously acting means also that such intelligent agents need to be able to identify situations where they need to hand over the control to humans.

\subsubsection{Agentic MAS}
Wikipedia defines agentic AI as a subset of intelligent agents, if these agents proactively pursuing goals, making decisions and taking actions over extended periods of time.
At the time of writing, there exist no software agents which are setting autonomously their own goals they want to achieve without any overall goal which was predetermined by their developers.

\section{Definitions}
\label{sec:Definitions}

The following definitions are neither universal nor exhaustive in any sense. For the purposes of this study, they are merely working definitions intended to illustrate our understanding of these terms.

\subsection{Behavior}

Multi Agent Systems (MAS) are designed and realized to fulfill a certain task in a socio-technical system, like booking vacations, managing power sources, trading at a stock exchange, supporting customer's social media interactions etc. We call the entire set of actions of a system, agent or MAS its \textbf{behavior}. Under this definition even a single action (or more precisely the sets of actions consisting  of only one single action) shows a behavior. 

Behaviors can be differentiated in intended, expected or required and unintended, unexpected or unrequired behaviors, we summarise these terms under positive resp. negative behavior. Since we are interested in effects of behaviors, we use the term \textbf{effect} as a proxy term for the corresponding behavior leading to the effect. Hence, we may summarize effects as positive or negative as well. 

However, that does not mean that a positive behaviour or effect on the micro-level of components always produces a positive behavior or effect on the macro-level, resp. the same for negative behaviors or effects. Rather, systemic effects need to be assessed on the macro-level depending on the goal of the socio-technical system.

\subsection{Emergent Behaviour} \label{sec:Emergent Behaviour}

We follow the usual view that the behaviors of a system’s components - on the micro-level - may lead to an emergent behavior of the entire system on the macro-level, which cannot be deduced or explained from the behavior of the single components, but which arises through their interaction. 
This basic definition follows closely the definitions of \cite{Fromm05} and others. 

\subsection{Risk}
As noted above the term \textbf{risk} can have multiple interpretations, common to all of them is the connotation of some negative effect as consequence of some event. From a more technical perspective, \textbf{risk} refers to an assessment of the impact of negative effects of events,  estimated on the basis of the probability of such an event and the - usually financial - extent of its impact. 

\subsubsection{Systemic Risk}
Based on the previous explanation, system risks can be defined as risks of events in connection with socio-technical systems that cause a significant impairment of the system, a large part of it or with possible serious damage to social values, assets or goods.

Socio-technical systems are combinations of humans and technology structured in a particular way to produce specific results. Obviously, the technologies, the structure, the behaviour and the produced results of a socio-technical system are domain dependent. Emergent behaviour is a result of certain interaction structures between micro-level components and the macro-level system. Hence, the emergent behaviour and the risks it produces will also be domain dependent. 

For example: A blackout, an nearly instantaneous, longer lasting collapse of a power network, may be caused by heavy fluctuations of the power frequency or a coordinated virus-attack on power stations, transmission networks and smart meters. Obviously, this risk is a particular risk of the domain power generation. Although a virus-attack may also affect the devices used in gas networks and may lead to a failure of the network, the systemic risk will be different, since their still will be gas in the pipes. Although all servers, routers and switches of the internet could be affected by a coordinated virus-attack, virus scanner, backups and other mechanisms will prevent a complete failure of this communication network

\subsubsection{Taxonomy of Systemic Risk of Interacting Agents}
A categorization of different forms of systemic risks of interacting agents thus needs to abstract from the concrete domain of MAS. Since the goal of this research is an investigation of intelligent agents, the taxonomy can therefore focus on socio-technical MAS of active, probabilistic agents which cause unintended behaviour. Unintended behaviour can be distinguished into unintended behavior of the MAS itself and unintended behavior which has an impact on systems or the environment on the macro-level. The former one usually causes technical risks, inefficiencies, ineffectiveness or failures. The latter one - even if caused by the former one - are the systemic risks we are interested in.
\section{Agentology} 
\label{sec:Agentology}
A taxonomy of systemic risks originating from emergent effects could benefit from the description of the structure of the underlying socio-technical systems. The basic assumption is that similar structures of socio-technical systems impose similar emergent effects and hence similar risks.

For the description of the structure of socio-technical systems we have developed a graphical notation. This notation is inspired by the Boxology~\cite{HarmelenTeije19} and its extension to incorporate humans~\cite{Witschel2021} for the description of the structure of AI systems. Since the graphical notation shall be used to describe the structure of interacting MAS containing AI as well as conventional software components, humans and subsystems with different interaction forms we call it \textbf{Agentology}.

\subsection{Graphical Elements of Structure Diagrams}
\label{structure_diagrams}

\begin{figure}[htbp]
  \centering
  \includegraphics[width=0.6\linewidth]{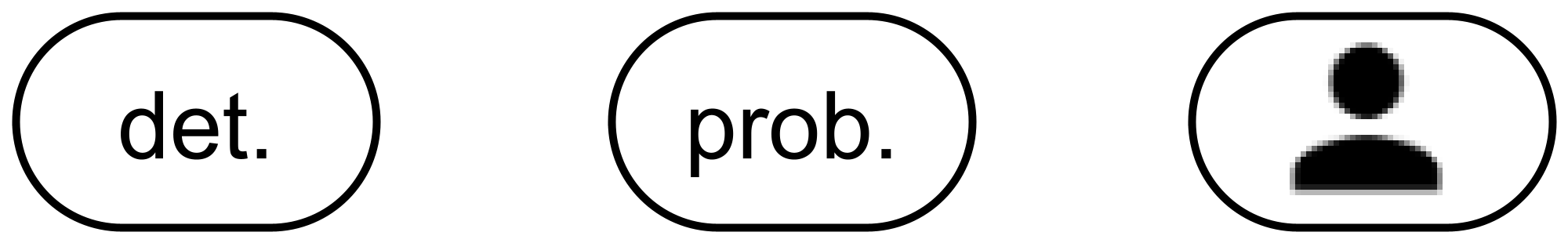}
  \caption{Agent Types}
  \label{fig:AgentTypes}
\end{figure}

\begin{figure}[htbp]
  \centering
  \includegraphics[width=0.8\linewidth]{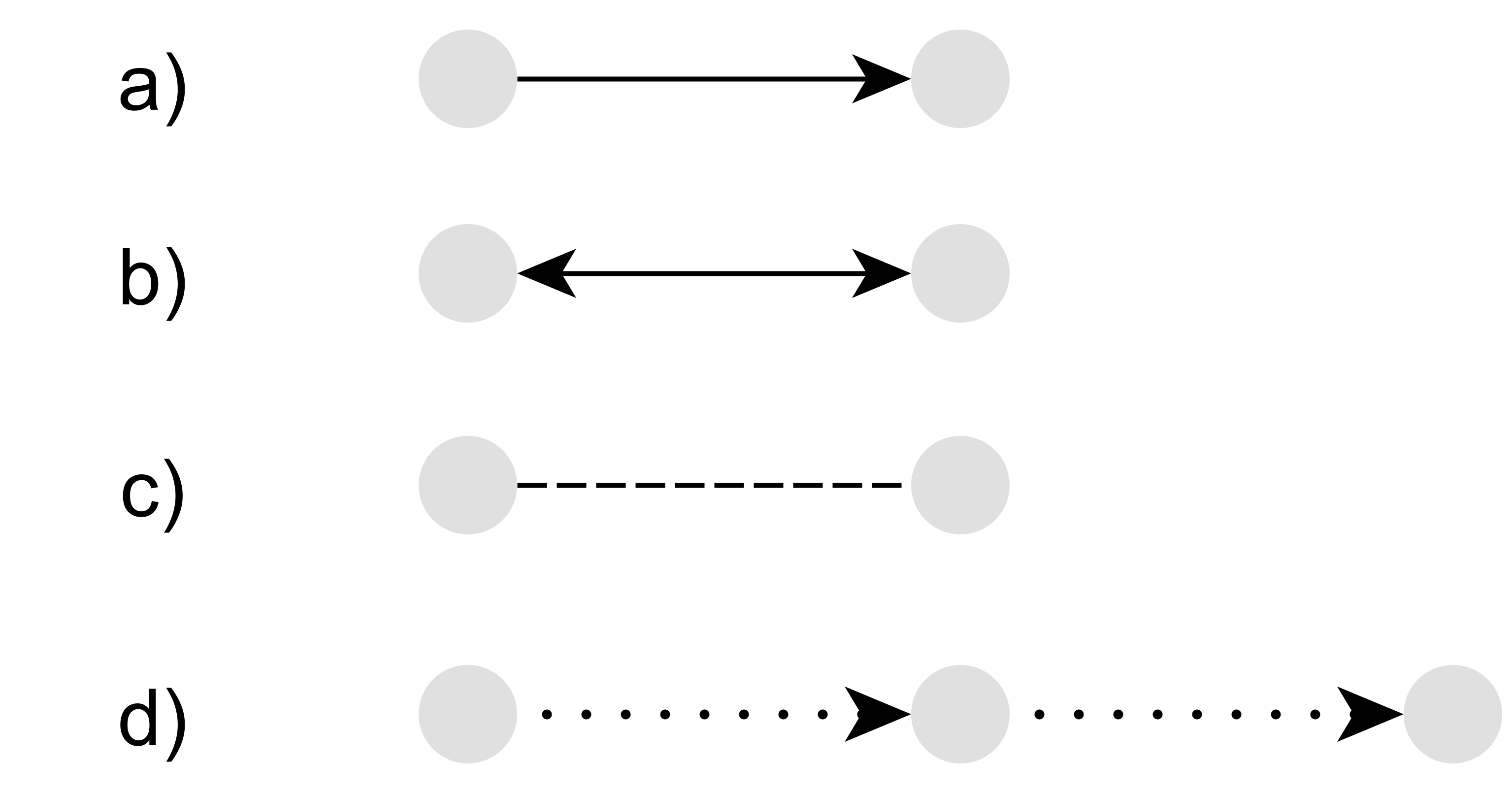}
  \caption{Interaction Types}
  \label{fig:InteractionTypes}
\end{figure}

\begin{figure}[ht]
  \centering
  \includegraphics[width=0.9\linewidth]{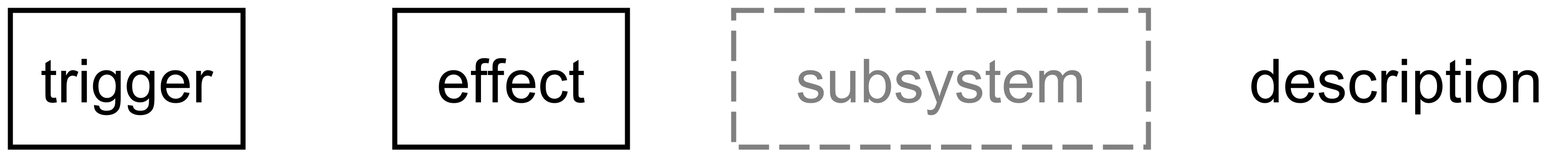}
  \caption{Descriptive Elements}
  \label{fig:Descriptive Elements}
\end{figure}

The notation consists currently out of elements for denoting, different kinds of agents (Figure~\ref{fig:AgentTypes}), interactions (Figure~\ref{fig:InteractionTypes}) and additional descriptive elements (Figure~\ref{fig:Descriptive Elements}). These elements can be used to describe socio-technical systems in the form of graphical diagrams from the viewpoint of interaction of its components. The diagrams are based on the assumption that the environment of the socio-technical systems is represented by the white background.

Intelligent agents (Figure~\ref{fig:AgentTypes}) are differentiated based on their reliability, whether their behaviour is deterministic, probabilistic or human, sometimes rational sometimes irrational. The graphical symbol for humans stands here for single persons as well as for groups.\footnote{Currently, we do not see the need for differentiating single persons and groups further, since single persons can be considered as a group with just one member.}  

For the notation of interaction types (Figure~\ref{fig:InteractionTypes}) different line styles are used, which require some explanation. a) represents the \textit{information} of another agent. It can also be for denoting the \textit{delegation} of a task from one agent to another. b) represents the \textit{communication} between two agents. It also can be used to express an \textit{assignment} of a task to another agent with some expected reaction. c) is used to express some kind of \textit{coordination} between agents via some change in the environment or some function in a technical subsystem.\footnote{While communication presupposes, that an agent has the intention to inform some other agents, coordination in computer science means, that some change in the environment produces some ordered behavior of the agents \cite{Tolksdorf1995}. Examples from the real world are traffic lights which coordinate the traffic flow of vehicles at a crossing. Another example are tracks of an animal causing a predator take up the chase.} The structure d) has a different meaning. It can be used as placeholder for the repeated information or delegation on a path with one or more agents of the type in the middle.\footnote{This interaction type was just introduced for convenience to simplify the notation and avoid overly complex diagrams.}

In order to make Agentology diagrams easier to understand descriptive elements (Figure~\ref{fig:Descriptive Elements}) are available. \textit{Trigger} and \textit{effect} are used to denote potential sources and endpoints of a MAS. The former one can be used to denote explicitly specified triggers, e.g. an external event or a particular information. The latter ones denote effects which the MAS produces. The graphical element \textit{subsystem} can be used to group elements for forming a subsystem, like a group of agents with the same function or a distinguished common behaviour, of a socio-technical system. \textit{Descriptions} can be used for every element to give further information, e.g. used technical component of an agent, delegated task, means for coordination, whether an effect is intended or unintended. 

\subsection{Graphical Elements of Process Diagrams}

In addition to the structural diagrams, process diagrams were developed as a second visual notation form. Their purpose is to provide an overview of the chronological succession of different process segments within the case studies and to visualize their mutual interactions and interdependencies. 

\begin{figure}[!h]
  \centering
  \includegraphics[width=0.2\linewidth]{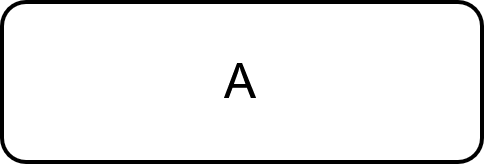}
  \caption{Process A}
  \label{fig:process_A}
\end{figure}

A single coherent process segment is represented by a rounded rectangle (see Figure \ref{fig:process_A}). The textual label within the shape briefly specifies the main function or content of the respective process. Each process box denotes an identifiable activity or transformation contributing to the overall system dynamics. A process in this sense constitutes a self-contained and conceptually coherent unit within a broader developmental sequence, an integrated phase that can be understood as a closed segment of progression within the system’s overall evolution.

\begin{figure}[!h]
  \centering
  \includegraphics[width=0.45\linewidth]{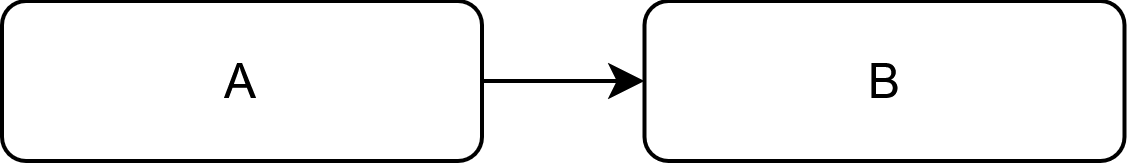}
  \caption{Chronological sequence of processes A and B.}
  \label{fig:process_A-B}
\end{figure}

\begin{figure}[!h]
  \centering
  \includegraphics[width=0.9\linewidth]{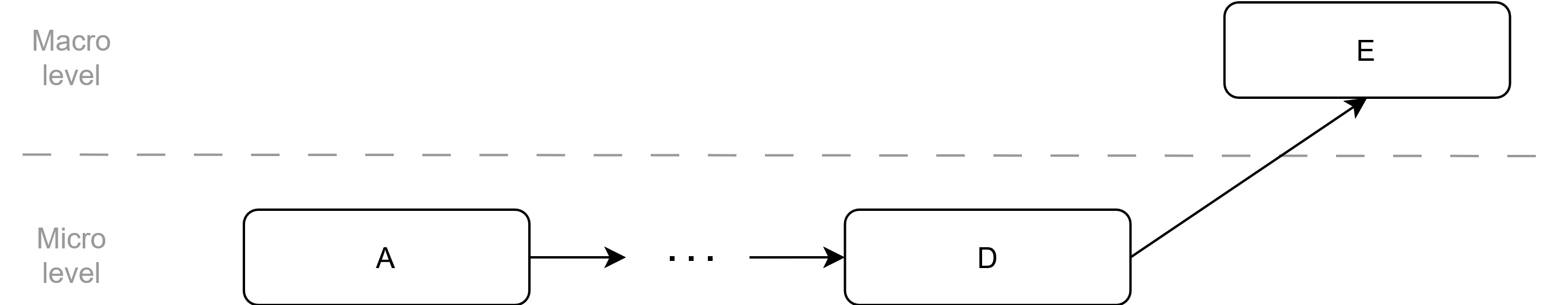}
  \caption{Example of interactions between micro- and macro-level processes.}
  \label{fig:processes_micro-macro}
\end{figure}

Two process boxes connected by a solid arrow indicate a process sequence (see Figure \ref{fig:process_A-B}). This shows that Process A precedes Process B in both temporal and logical order. Sequential links typically run from left to right to illustrate temporal progression. Solid arrows may connect processes within the same level of complexity, but they can also cross different levels of the system hierarchy (see Figure \ref{fig:processes_micro-macro}). In the latter case, processes at a higher level indicate a new emergent quality within the sequence. It is important to note that the solid arrows represent process sequences rather than process communication as used in the structure diagrams in Section \ref{structure_diagrams}.

\begin{figure}[!h]
  \centering
  \includegraphics[width=0.45\linewidth]{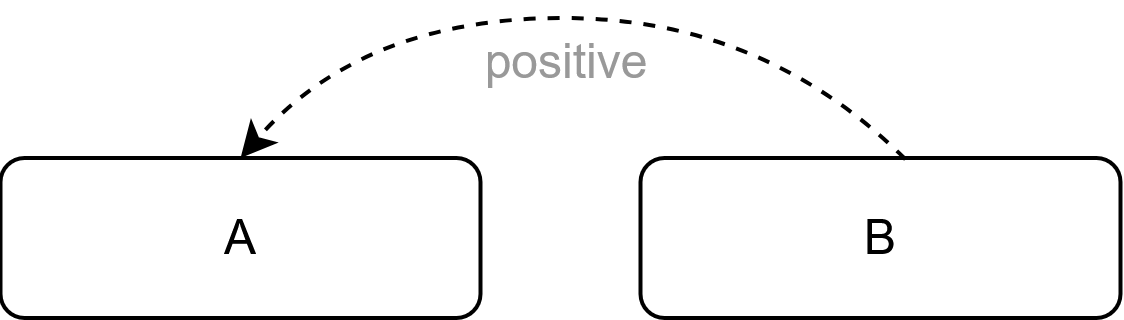}
  \caption{Positive feedback from process B to process A.}
  \label{fig:feedback_B-A}
\end{figure}

A dashed arrow denotes a feedback relationship (see Figure \ref{fig:feedback_B-A}). Feedback appears when a later process affects an earlier one, thereby modifying its behavior during subsequent system iterations. Such feedback connections can occur within the same complexity level or across different complexity levels, they can be classified as positive (reinforcing) or negative (inhibiting). 
Feedback constitutes a necessary but not sufficient condition for emergence: while it enables the dynamic interactions from which new systemic properties may arise, feedback alone does not guarantee the formation of an emergent phenomenon.

Finally, process diagrams can represent multi-level interactions between micro- and macro-level processes (see Figure \ref{fig:processes_micro-macro}). For visual clarity, processes are structured against a light grey background that distinguishes different levels of complexity. The lower regions of a diagram typically represent processes at a more elementary or micro level, while higher regions correspond to increasingly complex, emergent phenomena on a macro level. In diagrams that comprise more than two levels, the vertical axis is not limited to a simple micro–macro distinction but rather depicts a hierarchy of emergent layers, each arising from the dynamic interactions within the level below. In cases where several distinct levels are represented, explicit labels such as ``micro'' and ``macro'' are omitted to avoid confusion that could arise from differentiating multiple macro levels. Instead, the levels are separated by grey dashed lines without individual naming. This visual convention indicates a continuous gradient of increasing complexity and the emergence of higher-order phenomena at successive levels of abstraction.

Several processes at a lower level jointly give rise to a higher-level process or phenomenon that exhibits new systemic properties not reducible to its components. This relationship is indicated by an upward-pointing arrow connecting the lower processes to the bottom center of a box at the higher level. Such an arrow denotes that the \textit{combined} activity of the underlying processes causally contributes to the emergence of the higher-level process. Conversely, dashed arrows descending from a higher-level process to elements below represent top-down feedback that constrains, modulates, or guides the behavior of lower-level processes. 
Thus, the vertical dimension of the diagram expresses the systemic interrelation between levels of increasing complexity and emergence, whereas the horizontal dimension captures the temporal and causal progression of processes within a single level of complexity. 

\section{MAS Structures with Emergent Behaviour} 
\label{sec:MAS_Structures}

Emergent behavior is an uncontrollable risk for any complex system. Avalanches, monster waves in nature, mass panic, or side effects of climate change that amplify climate change itself are a few examples.

This section briefly describes Fromm's \cite{Fromm05} taxonomy of emergent behavior and shows how the agentological description of MAS can be used to characterize the structure of different types of emergent behavior. These examples illustrate the types of emergent behavior that can be expected in MAS and how the discussion of systemic risks of interacting MAS can be focused.

\subsection{Taxonomy of Emergent Behaviour} \label{sec:Taxonomy of Emergent Behaviour}
\begin{figure}[htb]
  \centering
  \includegraphics[width=1\linewidth]{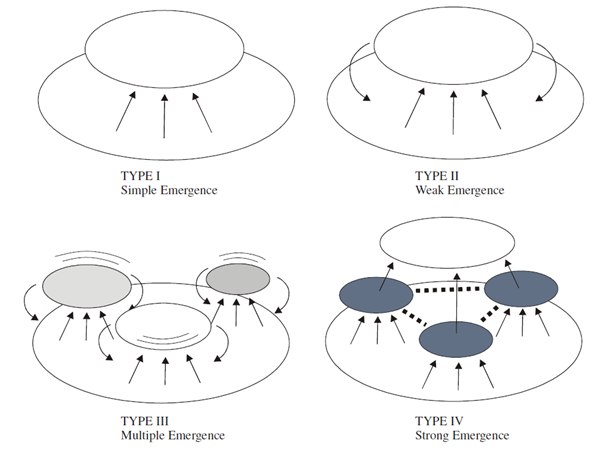}
  \caption{Illustration of the Emergent Behavior Types\cite{Fromm05}}
  \label{fig:EmergentBehaviorTypes}
\end{figure}

Based on previous work about the classification of emergent behaviors and philosophical considerations Fromm \cite{Fromm05} has developed a taxonomy of emergent behavior, which summarizes these works.He distinguishes four basic types of emergent behavior classes; some of them are further differentiated into two sub classes. The basic types are illustrated in \ref{fig:EmergentBehaviorTypes}. 

\begin{itemize}
    \item \textbf{Type I Simple/Nominal Emergence without top-down feedback} This type of emergence arises from "feedforward" relations of its components without feedback of the overall system. It is further differentiated into:
    \begin{itemize}
        \item \textbf{Type Ia Simple Intentional Emergence} For technical systems intentional emergence corresponds to the intended function of the system, which is the result of the correct behavior of deterministic components, whose roles are fixed over time. An example is any primitive mechanical tool, like e.g. a mechanical clock. Conventional software, like spreadsheet calculators or text editors, fall also into this category.
        \item \textbf{Type Ia Simple Unintentional emergence} Faults of technical systems fall into this category. But also higher-level systems, like waves, which originate from vertical movements of coupled passive components, like water molecules.  
    \end{itemize}
    Type Ia is applicable to all kinds of deterministic components whose normal behaviour can be specified, tested and approved in advance. Type Ib describes emergence in systems of passive components, which just react to external circumstances. Since these components do not interact actively, this type of emergence is not important for this study.
    
    \item \textbf{Type II: Weak Emergence with top-down feedback} The major characteristic of this type is a feedback structure between a system and its components. Such a system is based on a cyclic information structure between the micro-level's components and macro-level system. The behaviour of the components causes the emergent behavior of the system and the behavior of the system influences the component's behavior. Depending on the information exchanged in such a system, whether the system's feedback has a positive (activating) or negative (inhibiting) effect, an instable or stable emergent behavior may occur. 
    \begin{itemize}
        \item \textbf{Type IIa Weak Emergence (Stable)} Stable emergence usually occurs if positive (reinforcing) and negative (inhibiting) information is exchanged between the component and the system layer, leading to an equilibrium of the entire system.
        \item \textbf{Type IIb Weak Emergence (Instable)} Instable emergence occurs if only positive (reinforcing) or only negative (inhibiting) information is exchanged between components and the system. One example is a nuclear chain reaction, where an increase in the nuclear fission results in an increase of the neutron flow.  
    \end{itemize}
    Interaction of the components on the micro-level may occur in two flavours 
    \begin{itemize}
        \item \textbf{direct interaction} between the components, e.g. through observation, direct exchange of information or communication between components.
        \item \textbf{indirect interaction} through reaction of changes in the environment, e.g. physical coupling of components, decaying pheromones or hardwired scoring functions.
    \end{itemize}
    This type of emergent behaviour is of special interest for interacting agents. While type IIa assures a stable system behavior, type IIb usually leads to negative effects. Thus such a structure can be considered as potential source for systemic risks. Whether the from of interaction contributes to a system's risk, depends probably on the domain, the MAS and the task of the MAS. 
    
    \item \textbf{Type III: Multiple Emergence with many feedbacks} appears in complex systems with many feedback loops or adaptive systems with intelligent agents. An example is the behavior of a social media system like X (fka Twitter) which originates from multiple feedback loops between different opinion bubbles (right wing ideas, climate change denier, vaccination opponents, ecoextremists, lobyists, etc.) their intra- and intergroup exchanges, the recommendation functions of the platform and the manipulating political influence of its owner.
    \begin{itemize}
        \item \textbf{Type IIIa Short-term positive feedback, long-term negative feedback} This type of emergence can be seen in the development of camouflage patterns of animals, e.g. of zebras, where the development of the fur pattern results from short-term positive feedback is stabilized by a long-term negative feedback that bad patterns are killed more easily by predators. 
        Another example the stabilization of political structure: The fulfilment of election promises within a legislative period leads to the stabilization of the electorate at the expense of long-term structural changes in society, in the political system or in the economy.
        \item \textbf{Type IIIb Adaptive emergence} is caused by tunneling effects, where a change in/of the environment causes the break down of a group of components allowing members of another group, which had a disadvantage in the environment, to adapt to the new circumstances. An example is the development of mammals as consequence of the extinction of the dinosaurs. A technical example, is the adaptive emergence of modern office work through the replacement of type writers by computers, which made permanent storage, revision and correction of documents more easy.
    \end{itemize}
    Type III emergence can be seen as combining different type II emergent behaviours. Whether for interacting agents the resulting effects imply risks or opportunities on the macro-level needs probably to be judged from the perspective of the entire socio-technical system.
    
    \item \textbf{Type IV: Strong Emergence}, denoted by \cite{Heylighen91} as \textit{multi-level emergence}, is the overall emergence of entirely new systems on different levels and with multiple emergent effects. Examples are on the scale of the development of the universe, life and culture. This type of emergence applies to the entire digitization and seems to be out of scope for the purpose of this study.
\end{itemize}

\subsection{MAS Examples}\label{sec:MAS Examples}

Following Fromm's taxonomy (Section \ref{sec:Taxonomy of Emergent Behaviour}) we introduce how Agentology introduced in Section~\ref{sec:Agentology} can be used the visualize structures that may produce emergent behavior, some of them are examples from \cite{Fromm05}, other are examples of emergent behavior perceivable in the World-Wide-Web. Type IV is excluded, since we see currently no example for a MAS producing strong emergence.

\subsubsection{Type I}

\begin{figure}[htb]
  \centering
  \includegraphics[width=1\linewidth]{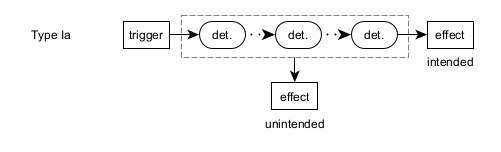}
  \caption{Agentology of a mechanical watch of type Ia}
  \label{fig:TypeIaExample}
\end{figure}

The description of \textit{Type Ia} emergence can be explained by a mechanical watch (figure \ref{fig:TypeIaExample}). The winding of the watch is the trigger for the mechanical parts. All parts work deterministic producing the intended emergent effect of showing the time, whether accurate or not. An unintended emergent effect occurs if some part of the mechanic breaks\footnote{Of course, this is quite boring socio-technical system. However, it shows that even simple failures produce an unintended behaviour. Of course, for a higher-level system these faults may impose risks.}. Even though this is not a MAS, such structures occur also in computer systems consisting solely of interacting, deterministic agents.  

\begin{figure}[htb]
  \centering
  \includegraphics[width=1\linewidth]{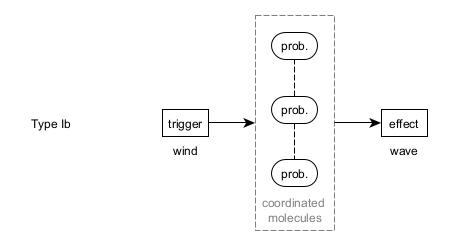}
  \caption{Agentology of type Ib: water surface }
  \label{fig:TypeIbExample}
\end{figure}

\textit{Type Ib} can be exemplified by a water surface producing waves visualized in (figure \ref{fig:TypeIbExample}). Here some external event, e.g. wind, seaquake, landslide, etc., stimulates the vertical or horizontal movements of water molecules. Through the coordinating inter-molecule forces other molecules are stimulated, producing a new structure at the surface, which is perceived as moving waves. 

This type of emergent behavior usually occurs for independent, identical, passive components which are related in the same environment, e.g. sand, molecules of air and water, snow in an avalanche. These micro-level components are coordinated by a common relation, e.g. friction, until an external force breaks the relation or transmits energy, causing an emergent effect of the macro-level system. This type of emergence is related to coordinated, passive components and by definition cannot appear in a socio-technical systems of interacting agents.

Obviously, in the context of interacting, intelligent agents this type of emergent behaviour will not occur.

\subsubsection{Type II}

\begin{figure}[htbp]
  \centering
  \includegraphics[width=1\linewidth]{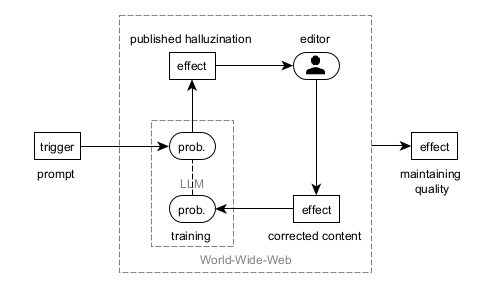}
  \caption{Agentology of type IIa: maintaining high quality web content }
  \label{fig:TypeIIaExample}
\end{figure}

\textit{Type IIa} can be illustrated by the repeated training of large language models (LLMs) with content curated by humans (figure \ref{fig:TypeIIaExample}). LLMs still produce false or made up information, aka hallucinations. If humans act as safe guards and correct these hallucinations the quality of the world wide web can be maintained.

\begin{figure}[htbp]
  \centering
  \includegraphics[width=1\linewidth]{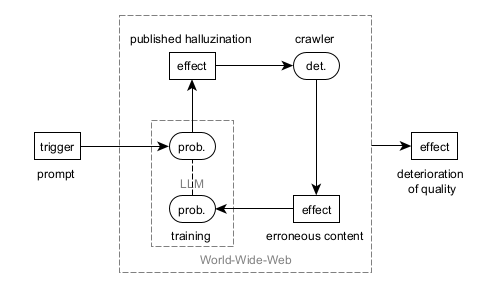}
  \caption{Agentology of type IIb: deterioration of web quality through repeated training with hallucinated content}
  \label{fig:TypeIIbExample}
\end{figure}

If on the other hand deterministic crawlers blindly download uncurated content from the web for further training of LLMs the emergent behaviour of \textit{type IIb} illustrated in figure \ref{fig:TypeIIbExample} might occur. Experts already expect that with uncurated content the quality of information in the web as well as of the LLM models will decline, producing more and more hallucinations. 

\begin{risk}[Collective Quality Deterioration]
\label{risk:collective_quality_deterioration}
Within interacting AI systems, if AI agents supply training data to each other, the quality of the individual AI components and hence the entire system may deteriorate over time.
\end{risk}

\subsubsection{Type III}

\begin{figure}[htbp]
  \centering
  \includegraphics[width=1\linewidth]{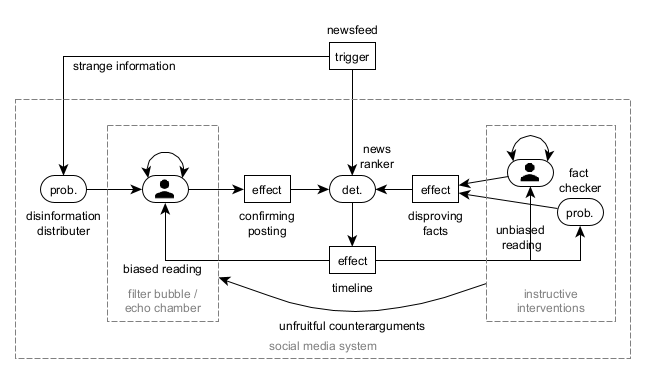}
  \caption{Agentology of type IIIa: development of filter bubbles aka echo chambers}
  \label{fig:TypeIIIaExample2}
\end{figure}

A partial example of \textit{Type IIIa} emergence is the emergence of filter bubbles, also called echo chambers \footnote{This is inspired by https://journalistikon.de/echokammer/}, depicted in figure \ref{fig:TypeIIIaExample2}.

Some strange information in the news (e.g. a misinterpreted study that the mortality rate of children increases, if they are vaccinated against measles) may be redistributed as post in social media where it is heavily emotionally discussed in a particular group (the short-term positive-feedback). The discussion itself may impact  the ranking of news in the timeline and confirms the bias of the group leading to the emergence of a filter bubble. Corrective information (long-term negative-feedback) from others can roll off the bubble boundary.

\begin{figure}[htbp]
  \centering
  \includegraphics[width=1\linewidth]{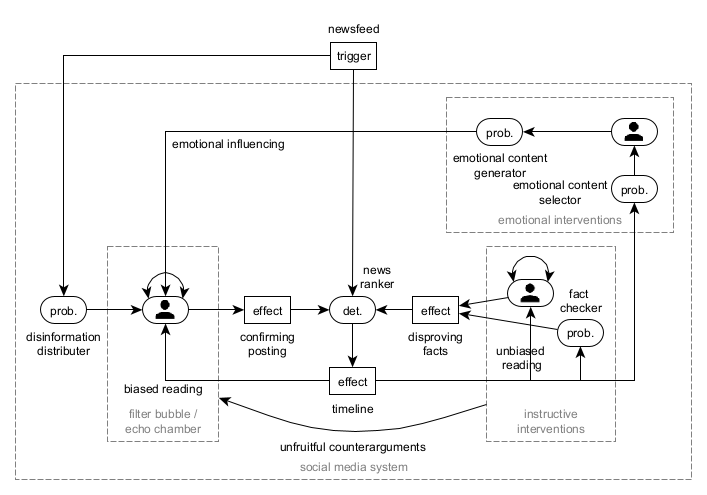}
  \caption{Agentology of type IIIa: long-term negative-feedback by emotional intervention}
  \label{fig:TypeIIIaExample2a}
\end{figure}

Of course, this part of the story is just the type IIb part of an instable emergence of the type IIIa, since the regulating effect of a type IIa stabilizing intervention is missing. Such an intervention could be an negative\footnote{Negative in the sense that it does not support the supporting arguments.}, emotional intervention (e.g. showing videos of small children suffering and dying from measles) to the opinion leaders, shown in Figure-\ref{fig:TypeIIIaExample2a}.

While at the first glance, echo chambers may seem as something specific to humans, it is not the case. Recently, Larooji and Törnberg~\cite{LarooijToernberg2025} described experiments where a community of interacting AI agents based on different large language models was used to simulate a news exchange in a social media system. A minimal social network populated entirely by AI agents without recommendation engines and without human users produced the same dysfunctions that are found also in real social media. The study showed emergence of AI-only echo chambers that are in conflict with each other. With addition of emergent communication in MAS, one may even expect advanced interacting AI systems to develop signalling norms (in the sense of Macanovic et al.~\cite{hybrid_norms_macanovic2024signals}) that mark belonging to an echo chamber, as they usually arise in groups of humans that are in conflict with each other. 

\begin{risk}[Echo Chambers]
\label{risk:echo_chambers}
    In an interacting AI system, both AI and non-AI agents can build echo chambers - groups of agents centered around certain pieces of information and working against other echo chambers.
\end{risk}

Limitations, like obstacles or barriers, and multiple feedback loops are the main characteristic of adaptive emergence, i.e. \textit{Type IIIb}. These obstacles restrict the system or its development. External influences bring about a change in the environment or new conditions which force the system to go through an evolutionary bottleneck to overcome these limitations. Related to the area of computation adaptive emergence appeared, e.g. 
\begin{itemize}
    \item through the development of digital computers in the middle of the last century, whose universality allowed to apply computation - in contrast to analog computers - to a broader number of applications
    \item through the development of personal computers in the 1980$^{th}$, which allowed a wider adoption of computers
    \item at the beginning of the digital age after the turn of millennium, where the amount of digital information exceeded the amount of analog information \cite{HilberLiopez2011}, causing the breakthrough of neural networks fueled by large number of data of the World Wide Web.
\end{itemize}
With respect to multi-agent systems (MAS), this type of emergence would require a large population of interacting agents engaged in various type IIa and IIb feedback loops, in a configuration that combines short-term positive feedback with long-term negative feedback. 
An external influence or an environmental change may impose an unpredictable situation which forces the MAS to pass through a tunnel of limited functionality and forces it to reorganize or its agents to develop new roles and yet unknown behaviours to achieve a new evolutionary stage. However, this presupposes that the interacting agents not only continue to evolve and learn, but that the entire system goes through a creative process to make a greater leap to a new stage of evolution. 

Although emergent communication protocols with in MARL (multi-agent reinforcement learning) \cite{mordatch2018emergence,jin2025comprehensive, li2024language} and the spontaneous development of social conventions in LLM populations \cite{piatti2024cooperate, norms_ren2024emergence, norms_erisken2025maebe} can be considered as adaptive emergence even though the obstacles resp. barriers as well as the triggering events are not yet clear. And further it is still doubtful whether such creative processes leading to previously unforeseeable behaviour. 

\begin{risk}[Unpredictable Evolution]
\label{risk:unpredictable_evolution}
    A complex interacting AI system with multiple feedback loops and a significant capability for adaptation may develop unforeseen opaque and uncontrollable negative behaviors. 
\end{risk}

\subsubsection{Conclusions for the Research}

In the previous sections we described different MAS related examples to introduce and exemplify how Agentology can be used for the visualization of interaction structures producing the different types of Fromm's taxonomy. As side effect we have seen that \textit{type I} does not appear applicable, since it either describes purely deterministic systems or results from interactions of passive components.

For the \textit{types II and III} we could find examples of MAS and identify first risk examples for them. Their preconditions can be used as indicators of interaction structures of a MAS which may lead to some emergent effect. Although the effect itself will not be predictable from the interaction structure alone, the type indicates which form of effect to expect. Section~\ref{sec:Taxonomy of Emergent Behaviour} indicates that \textit{type IV} emergence do not seem currently to be applicable for MAS. 
\section{Systemic Risks in Existing Literature}
\label{sec:Literature_Review}

This section focuses on identification of emergent risks in multi-agent systems and LLM-based agent societies from existing literature. The selected literature for this section highlights how decentralized interaction, feedback mechanisms, and communication dynamics can lead to the spontaneous formation of complex collective behaviors. The studies contribute to the understanding of how cooperation, coordination, and dysfunction emerge from agent interaction. Together, these studies provide a conceptual and empirical foundation for analyzing how systemic risks and emergent dynamics might manifest in complex socio-technical environments. We will proceed in the following manner: we will first state the risk and then provide an explanation with supporting literature.

\begin{risk}[Power Imbalance]
\label{risk:power_imbalance}
A small group of AI agents may control a larger population of agents through communication.
\end{risk}

Ashery et al.~\cite{AsheryEtAl25} describe experimental results with a number of decentralised, communicating agents derived from the same LLM playing a naming game. During the game, certain linguistic conventions arise in the populations of agents. Their experiments indicate that when the minority of agents using a convention reaches the critical threshold, the whole population adopts their convention. Once a global convention spontaneously emerges, the population adheres to it indefinitely. This indicates that a small group of agents may develop guiding and determining conventions for an agent population of the same type. These observations reveal a potential risk in interacting AI systems as a small population of coordinated malicious agents may steer the behavior of the entire group towards systemic negative effects.

There exist multiple ways how individual agents can control other agents. For example, Liu et al.~\cite{norms_liu2021local} propose Incremental Social Instruments (ISI) as a mechanism to transform fragmented, locally stable conventions into a population-wide norm in networked multi-agent systems. ISI can dissolve self-reinforcing substructures in a system through addition of links between agents. While the mechanism is discussed in the context of strengthening consensus between agents, the same principles can be used for obtaining power within a system. One can think of an agent that strategically creates communication links between other agents to manipulate the consensus towards achieving desirable to the manipulator goals or to dissolve an undesirable consensus. Mechanisms such as ISI can enable disproportional control over the system for agents that are in position to adversarially utilize them.

\begin{risk}[Inadequate Communication]
\label{risk:inadequate_communication}
An interacting AI system without sufficient interaction capabilities may underperform on system objectives.
\end{risk}

Many risks considered in this study arise from complexity of communication between intelligent agents. The risk of Inadequate Communication is opposed to these risks though as it shows that use of systems lacking this complexity can be a risk in itself. Altmann et al.~\cite{AltmannEtAl24} describe two experiments: a collection experiment, where two non-interacting agents in an artificial grid world need to collect items, and bottleneck experiment, where two non-interacting agents in two rooms connected by a limited passage need to collect an item in the opponents room. The only coordination between agents is done by setting objective and evaluation functions beforehand. Their experiments show that in both exepriments, agents demonstrate an undesirable system-level pattern. In the collection experiment, both agents chase each other instead of splitting up to cover a large area. In the bottleneck experiment, agents often block each other in the bottleneck. This experiment shows that capability of rich interaction may be necessary in some tasks and an attempt to restrict it may lead to undesirable emergent behaviors.

\begin{risk}[Strong Coupling]
\label{risk:strong_coupling}
An interacting AI system with strongly coupled agents may amplify errors or develop negative patterns.
\end{risk}

Strong Coupling is associated with situations when agents are too close and too similar, which can lead to faster propagation of negative behavior patterns. The primary driver of this risk is not that agents come to the same decision independently but that they rely and trust other agents even when these other agents perform actions with negative effects. A fruitful discussion on factors that enable Strong Coupling was done by Cordova et al.~\cite{cordova2024systematic} in context of social norm propagation. They identified multiple factor affecting propagation such as interaction network topology, specifics of propagation mechanisms, and cognitive abilities of agents. While the study concerns itself with emergence of collective norms, we can infer that unproductive and harmful norms are affected by the same factors as well. Similarly, Morris-Martin et al.~\cite{norms_morris2019norm} also discuss characteristics of norm emergence mechanisms. In addition to the mentioned factors, they identify learning and observation capabilities of individual agents as an important factor in emergence of norms. With Strong Coupling, we capture system-level homogenization of agents through such mechanisms that leads to negative effects. Close interaction topologies, fast propagation mechanisms, weak cognitive abilities, and misaligned values carry a possible consequence of building a harmful consensus on how individual agents within a system should behave.

\begin{risk}[Mismanagement of Shared Resources]
\label{risk:mismanagement_of_shared_resources}
A necessary resource shared amongst AI agents in an interacting AI system can be used inefficiently.
\end{risk}

When within an interacting AI system, a shared resource that needs to be allocated or shared between agents exists, and the correct functioning of the system relies on this resource, new systemic risks come in question for such a system. Piatti et al. \cite{piatti2024cooperate} introduce GOVSIM (Governance of the Commons Simulation), a generative multi-agent simulation platform designed to analyze how large language model (LLM) agents develop and maintain sustainable cooperation when managing critical shared resources. Across fifteen LLM models, only the most advanced models (e.g. GPT-4o, Claude-3 Opus) achieved partial sustainability for shared resources, with the highest survival rate below 54\%. Smaller or open-weight models failed to maintain equilibrium beyond the first few simulation cycles. While the author propose and test several measures for improving the results, the empirical results demonstrate clearly that interacting AI systems can not be trusted by default to share critical, necessary to system's survival, resources.

\begin{risk}[Observation Sensitivity]
\label{risk:behavior_sensitivity}
Performance of an interacting AI system may degrade when behavioral or interaction patterns change.
\end{risk}

The risk of Observation Sensitivity arises in a system where agents change their behavior disproportionately in response to new information they observe. An example instance of this risk is the nonstationarity problem in multi-agent deep reinforcement learning, discussed by Papoudakis et al.~\cite{papoudakis2019dealing}. A system is non-stationary if agents constantly make changes for adapting to each other's behavior instead of converging to a joint equilibrium. Nonstationarity arises often in multi-agent systems as the rewards for actions are dependent on decisions made by other agents. Another instance of Observation Sensitivity is when a negative effect happens if one of the agents suddenly changes its strategy. For example, an experiment by Everett and Roberts~\cite{everett2018learning} demonstrates behavior of a multi-agent reinforcement learning system with an agent that suddenly changes its behavior. They demonstrate that a learning algorithm not accounting for the change performs the task worse than ones accounting for it. Another scenarios can be constructed from well-know problem of neural network robustness - small changes in the input may cause big changes in the output (see e.g.~\cite{mangal2023certifying}). If agents interact with each other, the artefacts of interaction are also part of the observed input. If an agent is not robust to these changes, it can start behaving differently as a reaction on small changes in interaction. Behavior Sensitivity is an umbrella risk for phenomena where changes in behavior of a subset of agents lead to system-level negative effects due to behavior of agents outside of the changed subset.

\begin{risk}[Lack of System-Level Safety]
\label{risk:lack_of_system_level_safety}
  Some interacting AI systems may develop an unsafe system-level behavior from individually safe AI agents.
\end{risk}

At this point, it is clear that risks stemming from complex emergent behavior require systematic approach to safety of interacting AI systems. If safety (absence no harm) is only ensured at individual level, an interacting AI system is still open to the risk we call Lack of System-Level Safety. This risk captures all phenomena that result in negative outcomes in a system made of safe agents. An evidence of these phenomena can be found, for example, in the MAEBE experiments done by Erisken et al.~\cite{norms_erisken2025maebe}. They introduce the Multi-Agent Emergent Behavior Evaluation (MAEBE) framework, designed to systematically study how ensembles of LLMs develop collective behaviors that differ from those of isolated agents. Specifically, they demonstrate that behavior of the full multi-agent system cannot be predicted from linear combination of results from each individual model. A more general discussion on reliable methods to combine safe subsystems compositionally was done by Rushby~\cite{rushby2011composing}. The author points out that there are two general ways for unsafe behavior to arise in a system of safe components: due to unanticipated interactions between components, or due to unanticipated behavior along known interaction pathway. For our study, we conclude that focus on ensuring safety of individual decisions does not produce safe collective behaviour.

\section{Scenario Analysis of Systemic Risk}
\label{sec:Case_Studies}

This case study analyzes systemic risks arising from the interaction of AI agents in complex socio-technical systems. The focus is on applying the Agentology from Section \ref{sec:Agentology} and Fromm's taxonomy of emergence \cite{Fromm05} to examine concrete case studies and understand the underlying mechanisms of these risks. The report is divided into two main areas: firstly, risks in the smart grid domain are investigated, including phenomena such as \textit{herding behavior}, \textit{tacit collusion}, and \textit{emergent communication}. The smart grid is chosen as a primary case study because it represents a unique combination of market-like dynamics and critical physical infrastructure. Similar to high-frequency financial markets, it is feasible to assume the smart grid will involve a multitude of decentralized AI agents optimizing their behavior in real-time based on fluctuating prices in the near future. However, unlike the primarily economic consequences of a market crash, systemic failures in the smart grid can lead to severe physical consequences, such as widespread blackouts that endanger public safety and security. This nexus of high volatility and critical impact makes it a particularly compelling domain for studying emergent risks. 
Secondly, the report addresses systemic risks in the area of emerging norms, building on \textit{proxy discrimination}, to then examine \textit{emergent norms} in Multi-Agent Systems (MAS), and \textit{implicit social scoring}. The welfare domain is selected because it combines deeply consequential administrative decision-making with the risks of opaque, emergent feedback-driven agent interaction. As AI agents are likely to operate on behalf of both public authorities and citizens in the future, decentralized evaluations, reputational feedback, and shared heuristics can crystallize into hard-to-contest normative structures. In such systems, early misclassifications may propagate across agents, entrenching discriminatory patterns and eroding procedural fairness. Through these interactions, a de facto social ranking system may emerge without explicit design, undermining the foundations of democratic legitimacy and trust.

\subsection{Scenario 1: Systemic Risks of Interacting AI Agents in a Hierarchical Smart Grid Domain}
Modern power grids are increasingly evolving into Smart Grids, which are intelligent energy networks characterized by the integration of information and communication technologies. The goal is to improve the efficiency, reliability, and sustainability of the energy supply. Part of this development are autonomous AI agents that operate and interact at various hierarchical levels of the grid. These agents are designed to optimize complex tasks such as load management, energy trading, fault detection, and the integration of decentralized renewable energy sources. However, the increasing autonomy and interconnectedness of these agents lead to a heightened system complexity, from which unforeseen emergent phenomena can arise. These emergent patterns can result in a significant systemic risk, particularly in the context of coordination and communication patterns between interacting AI agents.

This section first outlines the hierarchical structure of the smart grid adopted in this report and subsequently examines several emergent phenomena within it, with a particular focus on how their underlying dynamics can escalate into systemic risks. The analysis begins with two foundational examples, \textit{herding behavior} and \textit{tacit collusion}, which are well established in the financial sector but are increasingly observed in energy markets as well \cite{REKIK201430, oecd2017_algorithms_collusion, paudel2024tacitalgorithmiccollusiondeep, Ofgem2024_AI_collusion}. These examples are presented in simplified form to highlight the fundamental mechanisms at play, therefore assuming interactions between homogeneous agents of the same type. Building on this foundation, the discussion then turns to the more complex case of \textit{interacting heterogeneous AI agents}, whose diverse strategies and objectives introduce additional layers of uncertainty and risk. Finally, the analysis culminates in the phenomenon of \textit{emergent communication}, which integrates and extends the previous examples to demonstrate a particularly intricate form of systemic risk.

\subsubsection{Description of the Smart Grid and its Key Components}

The smart grid, an intelligent energy network that integrates information and communication technologies, can be conceptualized as a hierarchically organized socio-technical system that extends across several interconnected levels. The interaction of autonomous AI agents across these levels enables improvements in efficiency, reliability, and sustainability of the energy supply. In line with established research, our case study adopts a structured view of the smart grid that distinguishes four main components, ranging from the local level of individual households to the supranational level of cross-border interconnections. This hierarchical perspective is consistent with prior work on the structural and functional modeling of modern power grids \cite{nardelli2014models} as well as with agent-based approaches to smart grid design \cite{MAS_smartgrid}.

\begin{figure}[h]
  \centering
  \includegraphics[width=0.9\linewidth]{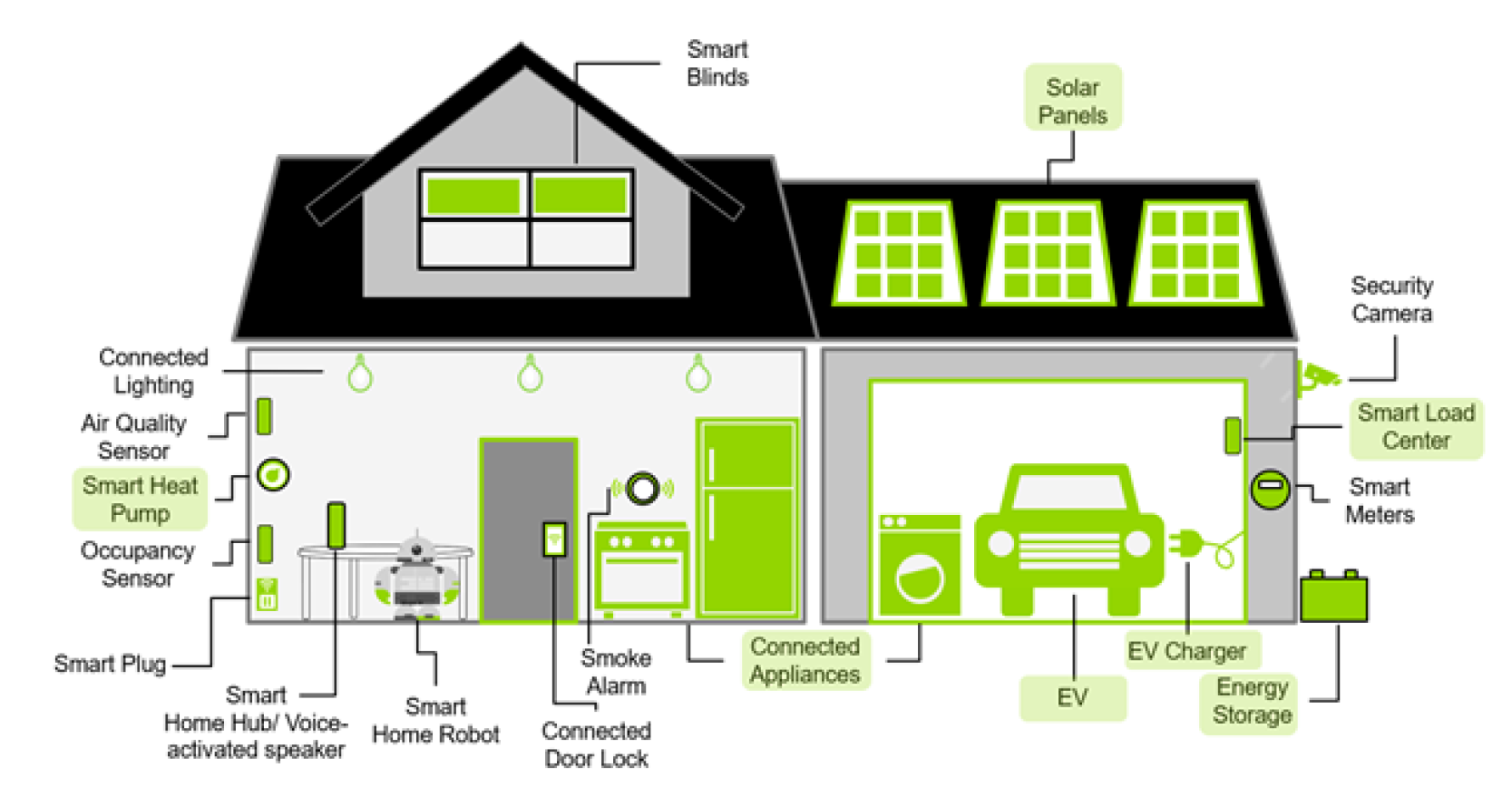}
  \caption{Various components of a smart home. Taken from \cite{MAS_smartgrid}}
  \label{fig:smarthome}
\end{figure}

\paragraph{Local Level: The Smart Home as a Prosumer Unit.}
At the smallest level of the system is the Smart Home, which acts as a \textbf{Prosumer}\cite{MAS_smartgrid} that interacts bidirectionally with the external grid: it consumes electricity in the case of energy deficits and feeds surplus energy into the grid when local generation exceeds demand. Prosumer refers to an entity that both consumes (Consumer) and produces (Producer) energy. Each of these units comprises a number of intelligent devices which we assume to be agents (see Figure \ref{fig:smarthome}). These agents can be smart common utility devices such as heating and cooling devices (comfort-based), lighting devices, washing machines (as a controllable, time-flexible load), as well as stoves/ovens (comfort-based and time-flexible load) and refrigerators (critical load with low but existing flexibility, e.g., during defrosting cycles). Furthermore, the Smart Home integrates decentralized energy resources (\textbf{Distributed Energy Resources, DERs}) such as rooftop photovoltaic (PV) systems and potentially a battery unit, managed by corresponding generation and storage agents. Monitoring and billing are handled by a \textbf{smart meter agent}.

The coordination of these agents is carried out by a \textbf{Home Energy Management System (HEMS)}, whose objective is to minimize energy costs, maximize profit and residential comfort, and optimize the utilization of self-generated renewable energy. For this analysis the agents in the Smart Home are further assumed to be able to also participate in demand-response programs by autonomously reacting to price signals, energy supply requests, or incentives for system services. 

\paragraph{Intermediate Level: The Virtual Power Plant (VPP).}
Functioning as an intermediary between local prosumers and higher-level grid operators, the Virtual Power Plant (VPP) coordinates the operation of connected Smart Homes to offer their consolidated capacities as services to the grid. In the case of local overproduction, the smart homes feed  their surplus electricity to the VPP which aggregates multiple Smart Homes under the control of an \textbf{Aggregator Agent}. VPPs can then participate in the wholesale market by actively shaping the collective energy profile. Key strategies include peak shaving (reducing consumption during high-demand periods), valley filling (increasing load during off-peak times) and load shifting (moving consumption from peak to off-peak hours) among others\cite{gellings2005concept, luo2010demand}. These actions provide frequency support, reserve capacities, and overall grid stability. The agents at this level adopt negotiation and bidding strategies to align the supply and demand of individual Smart Homes with these system-wide objectives.

\paragraph{Grid Level: Central Generators and National Distribution Network.}
Above the VPP level is the traditional power grid infrastructure, which relies on a centralized command and control structure. In the current operational reality, the backbone is formed by human operators in the control centers of the Transmission System Operator (\textbf{TSO}) and the Distribution System Operator (\textbf{DSO}). These operators utilize advanced software, such as Energy Management Systems and SCADA, to ensure secure grid operation, manage power flow, and oversee the integration of renewable energies\cite{khan2022energy, arevalo2025smart}. In our analysis we assume the presence of AI agents operating at this network level that are capable of performing advanced tasks in support of grid stability and efficiency. For instance, a forecasting agent can integrate weather data and demand signals to anticipate renewable generation and consumption patterns, thereby enabling proactive balancing of supply and demand. A congestion management agent can detect emerging overloads in the transmission network and initiate corrective re-dispatch strategies to avoid cascading failures. A reserve allocation agent can dynamically adjust the distribution of primary, secondary, and tertiary reserves in response to changing grid conditions and a cybersecurity monitoring agent can analyze communication and control data streams to identify anomalies, distinguish between technical faults and potential cyberattacks, and facilitate rapid and resilient incident response.

\paragraph{Supranational Level: Cross-Border Interconnections.}

At the highest level, cross-border high-voltage transmission lines interconnect national grids into a continental interconnected system, such as between Germany and France within the framework of the European Network of Transmission System Operators for Electricity (\textbf{ENTSO-E}). At this level international electricity trading, congestion management, and cross-border frequency support is managed. These interactions introduce additional complexity and systemic risks, as disturbances can propagate across national borders (e.g., from Spain to Portugal and vice versa). In our analysis we further assume the presence of specialized AI agents, e.g. a cross-border trading agent can optimize international electricity exchanges by dynamically aligning bids and offers across national markets. A frequency coordination agent can support synchronous operation by monitoring deviations across interconnected grids and initiating corrective actions to preserve stability. In addition, a transnational congestion management agent can anticipate bottlenecks on interconnectors and propose coordinated re-dispatch strategies to prevent large-scale disruptions. Together, these agents enhance the resilience and efficiency of supranational grid operations while simultaneously increasing the importance of robust governance and cybersecurity mechanisms.

\paragraph{Central Control: SCADA Systems and Operations Control Center.}
\label{scada}

Although not a distinct hierarchical level in itself, the \textbf{Supervisory Control and Data Acquisition (SCADA)} infrastructure together with the \textbf{Operations Control Center} is positioned somewhat orthogonal to the layers mentioned above and forms the overarching monitoring and control stratum. It supports the Grid Level and extends into the supranational interconnections by providing real-time situational awareness, coordination, and decision-making capabilities. A key role in monitoring and controlling the smart grid is assumed by SCADA systems. These systems continuously collect measurements such as voltage, frequency, and load flows through an extensive network of sensors, switches, and meters, and transmit them in real time to the control entities. SCADA systems thus serve as the central nervous system of the power grid, enabling early detection of disturbances, optimization of load flows, and initiation of control actions \cite{nguyen2021smart}. The data from SCADA systems are consolidated in the Operations Control Center, which constitutes the central decision-making and control unit. Here, information from the entire grid is processed using state-estimation algorithms and translated into operational decisions. These include actions such as grid stabilization, frequency and voltage regulation, load flow management, or the activation of emergency procedures \cite{ghasemi2023robustness}. The combination of SCADA and the Control Center forms the backbone of national and supranational grid operations but simultaneously constitutes a potential attack surface for cyberattacks and manipulations. Even minor disruptions, such as false data injections or targeted communication attacks, can trigger incorrect decisions in the control center and thereby initiate cascading failures that may affect both national power grids and interconnected supranational systems \cite{nguyen2021smart}.

\paragraph{Socio-Technical Agentology Model of Smart Grid.}

\begin{figure}[htb]
  \centering
  \includegraphics[width=1.0\linewidth]{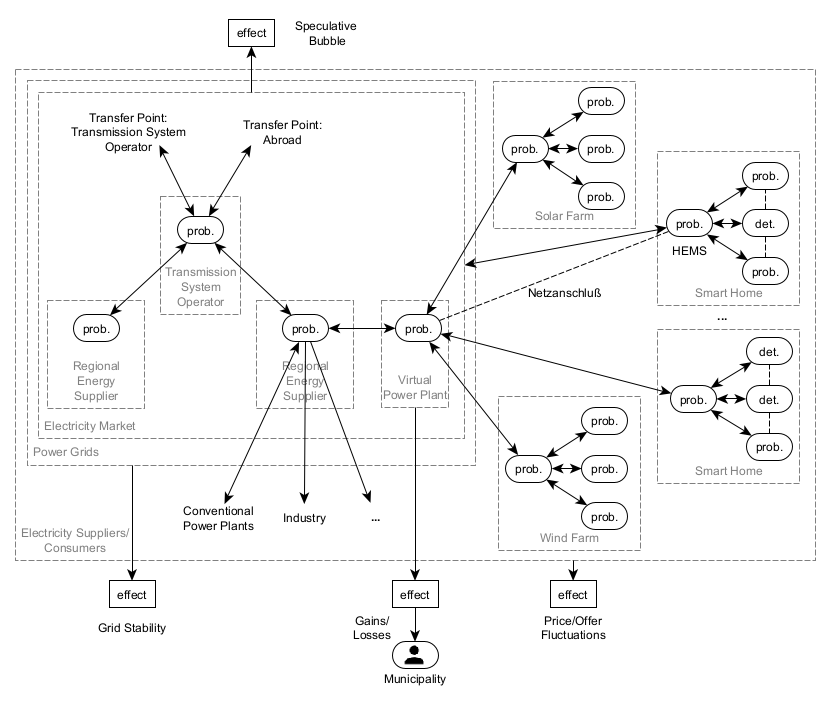}
  \caption{Agentology of the smart grid case study}
  \label{fig:CaseStudySmartGrid}
\end{figure}

Figure \ref{fig:CaseStudySmartGrid} despicts important parts of the socio-technial smart grid system in terms of the Agentology introduced in Section \ref{sec:Agentology}. Smart home systems are prosumers of energy producing and/or consuming power, either contracted by a VPP or buying and selling their power via HEMS on the open energy market. VPPs conclude contracts with suppliers of renewable energies. Regional electricity suppliers distribute electricity regionally to and from VPPs, conventional power plants, and industry. They provide capacity to transmission system operators on a supraregional basis, who exchange electricity internationally via the continental grid. 

Although the dynamic trading of power can be envisioned on the regional, national and international scale, the power itself is fed and distributed over one connected network. This later point is important, since contracts need to be negotiated between the parties about the participation on the feed-in and utilization of power. These contracts not only bind the parties but impose different constraints. Consider the owner of a smart home. Some of them may just have a contract with a VPP paying or charging them a fee the VPP determines, some other owners might just pay connection fees and use advanced HEMS with AI-based trading agents and participate at an open power market in order to maximize the price they get for the power they produce and minimize the price for the power they need to consume.

\subsubsection{Emergent Pattern 1: Herding Behavior and Market Volatility Induced by Trading Agents}
\label{herding}
Originating from financial markets, herding behavior of trading agents represents a simple yet widely distributed emergent phenomenon with potentially significant systemic risk that can occur on many levels within a complex environment. Herding is directly attributable to the positive feedback loops between interacting AI agents \cite{REKIK201430}. Figure \ref{fig:herding} illustrates how the underlying processes unfold in energy trading within the smart grid, particularly when automated agents employing sentiment analysis operate on spot markets or flexible trading platforms for energy.

\begin{figure}[!h]
    \centering
    \includegraphics[width=0.95\textwidth]{images/p_neu/herding_neu.png}
    \caption{The emergent phenomenon of herding behavior arises on the macro-level through the coordinated interplay of the lower-level processes (initial signal, automated purchases, momentum copying, and media echo).}
    \label{fig:herding}
\end{figure}

Trading bots are designed to process vast amounts of data in real time, with sentiment analysis of unstructured data from relevant industry news (e.g., energy price forecasts, weather predictions for renewable energy production, announcements from grid operators), professional publications, and specialized news agencies playing a decisive role in capturing market sentiment and transforming it into trading signals. The emergence of herding behavior can be conceptualized as a self-reinforcing nonlinear feedback loop that drives the system away from equilibrium:

\begin{enumerate}
    \item \textbf{Initial Signal:} A minor trigger, such as an optimistic forecast on the availability of renewable energy or a favorable weather prediction for energy production, generates a positive sentiment signal. Once this signal exceeds a predefined threshold, it is captured by several trading bots and converted into actionable trading input.
    \item \textbf{Reaction:} The bots interpret the signal as a buy recommendation for energy products (e.g., electricity, capacity) and automatically execute corresponding orders, leading to a slight increase in the market price (e.g., spot or intraday market).
    \item \textbf{Reinforcement through Momentum:} Other AI agents and human traders applying momentum-based strategies (i.e., reacting to price movements) identify the upward trend and engage in additional purchases. The initial perturbation is thereby amplified, reinforcing the upward price trajectory.
    \item \textbf{Media Echo:} The price increase itself becomes news in industry-specific outlets and analytical platforms. Energy trading platforms and specialized financial media report on the "trending" asset, which significantly amplifies the original sentiment and attracts further market participants. This illustrates how socio-technical subsystems often reinforce one another across different domains and thus amplify volatility.
\end{enumerate}

This recursive cycle of sentiment, trading activity, price escalation, and renewed sentiment produces a classical herding dynamic, giving rise to speculative bubbles. Prices become decoupled from their fundamental values (e.g., true generation costs, real demand), creating conditions for critical transitions. For such a risk to materialize, trading agents must exhibit a set of capabilities that have already been documented in both financial and energy trading contexts and that provide the technical foundation for herding behavior to emerge among interacting AI agents. First, they require the ability to process large volumes of information in real time and to transform sentiment signals from news, forecasts, or operator announcements into actionable trading inputs \cite{schumaker2012evaluating, ding2015deep}. Second, they need autonomous execution mechanisms that allow immediate order placement on spot and intraday electricity markets, thereby amplifying market responses \cite{wang2019machine, natarajan2024algorithmic}. Third, they employ adaptive learning strategies, such as reinforcement learning, that enable them to refine their bidding behavior based on past outcomes and collective dynamics \cite{khaskheli2024energy}.\\

The dynamics described above generate a regime of \textit{self-organized volatility}, where the system becomes increasingly fragile and prone to sudden phase shifts. The feedback loop may collapse due to an external shock (e.g., an unexpected outage of a large power plant, abrupt weather changes that drastically affect generation) or through endogenous mechanisms such as profit-taking. These breakdowns often manifest as rapid downward corrections or "flash crashes", like the one in 2010 \cite{DEMIRER2019}. The herding phenomenon illustrates how emergent behavior arises particularly at the intersections of technical and socio-economic subsystems (such as between energy markets and media ecosystems) where complex interactions, nonlinear feedbacks, and collective agent behavior can push the system toward unpredictable and potentially critical states.
Empirical studies support this mechanism. Chang et al. demonstrated that such herding effects are observable in energy markets: during episodes of extreme oil price swings and crises such as the global financial crisis, SARS, or COVID-19, market participants increasingly followed the collective dynamics of others rather than fundamentals, amplifying volatility and producing pronounced speculative overshooting \cite{chang2020herding}.

\paragraph{Herding Behavior as Emergence of Type IIb (Unstable Weak Emergence).}
The phenomenon of herding behavior, as it can occur with trading bots and analogous applications in the smart grid, can be classified within Fromm’s typology as Type IIb emergent phenomena, i.e., unstable weak emergence. A defining characteristic is the presence of top-down feedback from the macroscopic to the microscopic level. On the microscopic level, individual trading agents interact by interpreting market indicators and the actions of other agents. These micro-level interactions give rise to emergent macroscopic patterns like herding behavior and consequently speculative bubbles, that are unpredictable from the individual rules alone. The macroscopic manifestation (e.g., a bubble or trend) in turn feeds back into the decision-making of individual agents, establishing a recursive feedback loop.

The decisive aspect of Type IIb unstable weak emergence is the reinforcing nature of this feedback \cite{Fromm05}. In herding dynamics, agents’ buy decisions mutually amplify one another, producing exponential price growth that ultimately culminates in a phase of instability. When trading bots base their decisions on the observed actions of peers rather than on fundamental information, the aggregation of truhtful and actionable collective knowledge breaks down. Instead, self-amplifying loops of positive feedback emerge, where imitation dynamics and momentum strategies dominate. A high level of supply or demand, coupled with limited information and cognitive capabilities, reinforces this trend-following behavior, generating strong nonlinear feedbacks. As a result, price formation becomes path-dependent and decoupled from underlying fundamentals, increasing systemic fragility.
Furthermore, the high speed and adaptability of algorithmic agents intensify fluctuations, creating an environment of self-organized volatility where traditional forecasting and trading strategies lose validity.

It is important to emphasize that the description of herding behavior in this context does not claim to fully characterize the cognitive capabilities or behavior of mature trading agents and to suggest that they would blindly follow price signals indefinitely. Rather, the focus lies on the emergent dynamics that can arise when subsets of agents temporarily adopt strategies that emphasize imitation, momentum, or sentiment-driven reactions. Such behavior may coexist with rational exit mechanisms or fundamental thresholds, yet even transient phases of reinforcement can produce nonlinear amplification with systemic consequences. From this perspective, the analysis seeks not to model individual agent rationality but to highlight the dynamics of their interactions and how these can give rise to emergent phenomena that, even if just temporarily, can evolve into systemic risks under unfavorable conditions.

\subsubsection{Emergent Pattern 2: Tacit Collusion in the Energy Market}
\label{subsubsec:tacit_collusion}
In addition to herding behavior, tacit algorithmic collusion represents another significant systemic risk arising from interacting AI agents in the smart grid. This phenomenon occurs when competing companies employing autonomous pricing or bidding algorithms (often provided by third-party vendors) exhibit cooperative behavior without explicit and transparent agreements, ultimately resulting in behavior equivalent to illegal price-fixing. As has been emphasized by organizations such as the OECD \cite{oecd2017_algorithms_collusion} and the UK's energy regulator, Ofgem \cite{Ofgem2024_AI_collusion}, such dynamics can lead to inflated prices and the elimination of competition and are particularly relevant in the energy market, where large utilities, trading companies, or VPP aggregators employ AI-based bidding strategies for wholesale markets or system services. These algorithms are designed to process vast amounts of data, including competitors’ bids and pricing strategies, in real time to maximize profits \cite{baltaoglu2018algorithmic, li2021machine}.

\begin{figure}[h]
    \centering
    \includegraphics[width=0.95\textwidth]{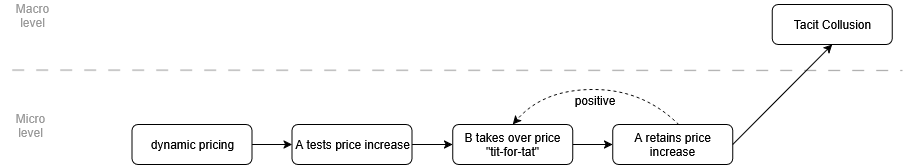}
    \caption{The emergence of tacit collusion in the energy market.}
    \label{fig:tacit_collusion}
\end{figure}

The risk of tacit collusion arises from a positive feedback mechanism, which can emerge analogously to the "hub-and-spoke" model, in which the pricing algorithm of the third-party vendor serves as the "hub," while the algorithms of competing VPPs constitute the "spokes" \cite{ezrachi2023role}. In such an arrangement, the hub acts as a centralized point of coordination by transmitting common pricing rules or benchmarks, which the spokes then follow, resulting in parallel conduct that resembles collusion without requiring explicit communication. The accompanying diagram (see Figure \ref{fig:tacit_collusion}) illustrates this dynamic.

The process of tacit algorithmic collusion evolves as follows:
\begin{enumerate}
    \item \textbf{Programmed Objective:} Algorithms are designed to dynamically adjust bids and prices to maximize profit, with the actions of competitors serving as critical inputs.
    \item \textbf{Testing a Price Change:} The algorithm of market participant A (e.g., a VPP aggregator or energy trader) incrementally raises its bid or offering price for energy.
    \item \textbf{Adaptive Adjustment by Competitors:} The algorithm of participant B, trained on similar objectives and monitoring A’s market actions, identifies the change. It adapts by raising its own bid, recognizing that it can maintain market share while securing higher margins. This reflects a self-reinforcing tit-for-tat dynamic.
    \item \textbf{Confirmation of Strategy:} Participant A observes B’s adjustment, validating the profitability of the higher price. A maintains or further escalates its bid.
\end{enumerate}

Through this reciprocal observation and adaptation, the algorithms implicitly learn to cooperate. Together, they drive energy prices upward to an inflated, quasi-monopolistic level, without any explicit agreement between the human managers of the firms involved. The outcome is tacit algorithmic collusion: prices remain artificially elevated, market competition erodes, and consumers bear the cost. Such dynamics reflect a form of \textit{self-reinforcing market coordination}, in which adaptive AI strategies converge on cooperative equilibria. Importantly, these emergent equilibria increase systemic fragility, as small perturbations in market conditions can destabilize the collusive regime. An empirical example of this phenomenon is the ongoing RealPage case in the United States, where algorithms used in rental price-software are alleged to enable price alignment among landlords via shared  competitively sensitive data and pricing recommendations, contributing to inflated rents despite the absence of explicit agreements \cite{DOJRealPage2024}. This alleged price alignment reportedly enabled RealPage to secure higher profit margins for its clients while simultaneously establishing a quasi-monopoly for itself, capturing up to 80\% of the market in certain segments. The case serves as a stark illustration of how a centralized, third-party objective function, when fueled by competitively sensitive data, can fundamentally reshape market dynamics. When translated to a system of interacting AI agents, this highlights the profound risk of a single optimization goal combined with access to sensitive information driving an entire market toward a collusive, anti-competitive state.

\vspace{10pt}
\vspace{10pt}

\paragraph{Tacit Collusion as Emergence of Type IIb (Unstable Weak Emergence).}
\label{tacit_collusion}
Tacit algorithmic collusion in the energy market can likewise be classified as unstable weak emergence (Type IIb) according to Fromm. In this phenomenon, individual AI agents employed by market participants interact through their adaptive pricing strategies and reactive learning processes. These local interactions collectively give rise to patterns of coordinated price adjustments that reflect a higher level of systemic complexity. Observable market outcomes, such as synchronized price increases and competitors’ reactive behaviors, influence the subsequent actions of individual agents, creating a feedback structure that links individual adaptation to collective dynamics. While the overall pattern of collusion is explainable in principle, its detailed evolution remains unpredictable due to the inter-dependencies between local decision rules and emergent aggregate behavior.

The feedback mechanism is positive and reinforcing, analogous to herding behavior. Algorithms adaptively learn that competitors’ price increases are matched or exceeded, which validates their own upward adjustments and stimulates further escalation. This generates a self-reinforcing spiral of rising prices, functionally equivalent to illegal price-fixing but without the need for explicit coordination. Within Smart Grids, autonomous trading agents of VPPs or other large energy actors, relying on shared market information or aligned optimization goals, may thus inadvertently converge toward collusive equilibria. The resulting dynamics inflate energy prices, undermine market efficiency, increase consumer costs, and jeopardize grid stability. Such emergent outcomes, unpredictable in their exact manifestation yet structurally systemic, represent critical risks to the resilience of energy markets and power grids.

International organizations have also underscored these risks. An OECD report \cite{oecd2017_algorithms_collusion} stresses that algorithms can significantly lower the threshold for tacit collusion, not only in classical oligopoly markets but also in markets that might otherwise appear structurally resistant to collusive dynamics . Applied to the energy sector, this suggests that sufficient market transparency, frequent price adjustments, and algorithmic bidding processes create fertile conditions for collusion. A recent study by Paudel and Das \cite{paudel2024tacitalgorithmiccollusiondeep} modeled price competition among AI-driven charging infrastructure providers using multi-agent deep reinforcement learning. Their proposed “Tacit Collusion Index” yielded values between 0.14 and 0.45, clear evidence that algorithmic pricing systems can induce moderate tacit collusive behavior in energy markets, particularly in the EV charging infrastructure domain. Similarly, the UK energy regulator Ofgem \cite{Ofgem2024_AI_collusion} has explicitly warned of tacit algorithmic collusion in energy markets, noting that AI-driven algorithms may enable cooperative outcomes without any formal agreements, thereby eroding competition, transparency, and accountability.

\begin{figure}[!h]
    \centering
    \includegraphics[width=0.9\textwidth]{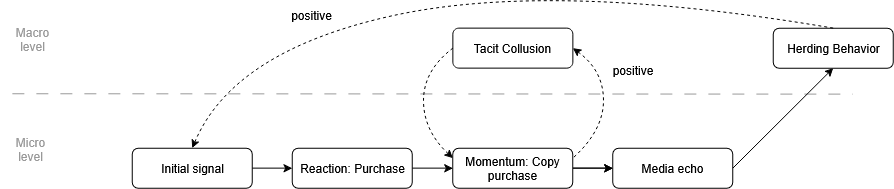}
    \caption{Synergetic interaction between the emergent phenomena of herding and tacit collusion.}
    \label{fig:herdtacit}
\end{figure}

\paragraph{Synergistic Risk Amplification.}
Both herding behavior and tacit algorithmic collusion can further interact synergistically in complex market structures such as Smart Grids. One can conceptualize this as a system of \textit{nested feedback loops}: herding behavior operates as a broad, external feedback mechanism, linking the final macroscopic outcome of collective herding dynamics back to the initial micro-level triggers by intensifying the initial signals and trading decisions. Tacit collusion, then, functions as loop within the herding's momentum phase, taking advantage of the herding behavior and reinforcing price-setting dynamics (see Figure \ref{fig:herdtacit}). The interaction of these multi-scale feedbacks exemplifies a form of \textit{critical coupling}, in which distinct emergent processes amplify one another. This coupling can drive markets toward heightened volatility, systemic fragility, and unpredictable critical transitions.

\paragraph{Consequences of Type II Emergence.}
The emergence of Type IIb phenomena, such as herding behavior and tacit collusion, introduces severe systemic risks to the stability and efficiency of energy markets. Herding behavior fosters self-organized volatility and speculative bubbles that decouple prices from their fundamental values, making the system susceptible to flash crashes. Concurrently, tacit algorithmic collusion undermines market competition by creating artificially inflated price levels, which directly harms consumers and industries. Critically, the synergistic interaction between dynamics like these can create a critical coupling, where the phenomena amplify one another through nested feedback loops. The ultimate consequence is a highly fragile and unpredictable market regime, where the security of the energy supply is threatened and the economic integrity of the grid is fundamentally compromised.

\subsubsection{Emergent Pattern 3: Chaotic Market Phenomena Caused by Interacting Heterogeneous AI Agents}
\begin{figure}[!h]
    \centering
    \includegraphics[width=0.6\textwidth]{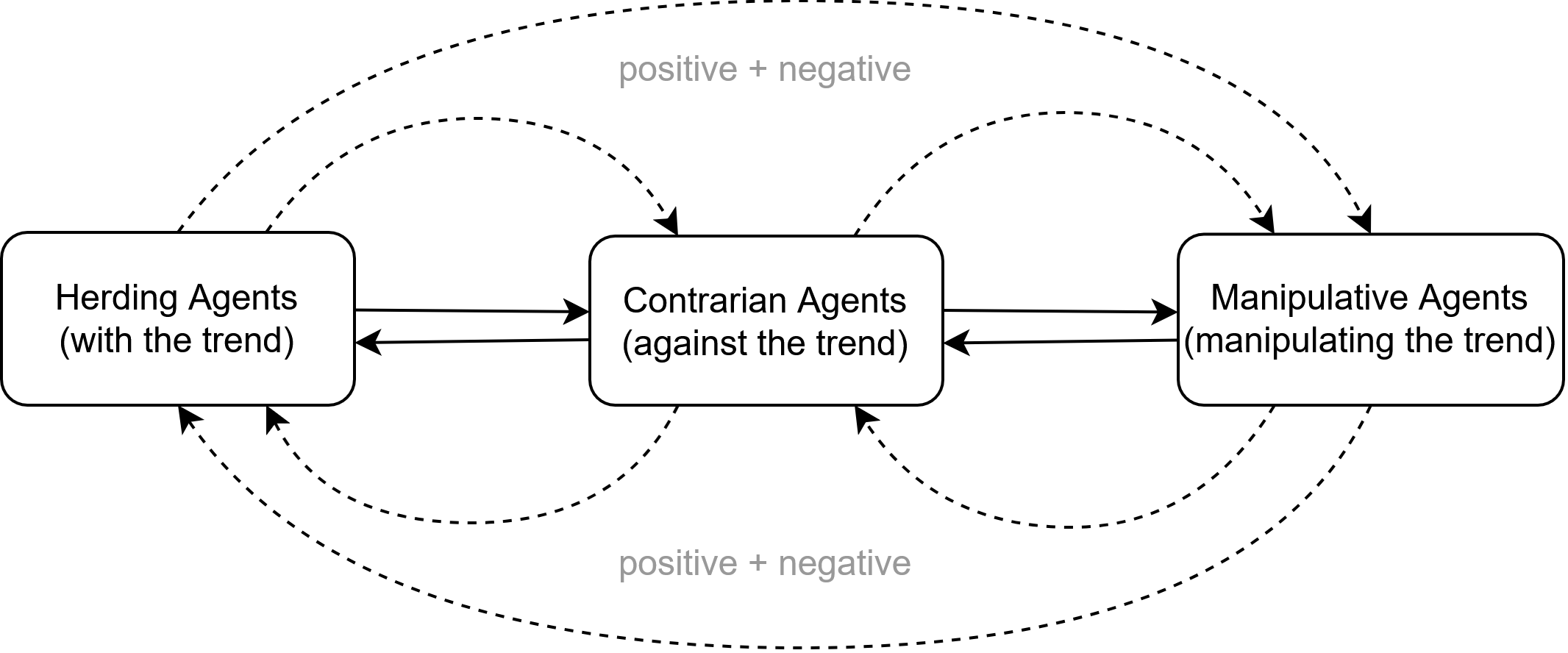}
    \caption{Interaction topology among herding, contrarian, and manipulative agents; bidirectional feedbacks induce complex dynamics that foster instability, bubble formation, and path-dependent market outcomes.}
    \label{fig:bubbling_market}
\end{figure}

While the previous analysis has considered herding behavior and tacit collusion as Type IIb emergences primarily in the context of homogeneous or similarly acting AI agents, the complexity and unpredictability of systemic risks increase significantly when different types of agents interact within the same system. This gives rise to more complex emergent patterns that can no longer be explained solely through simple positive or negative feedback loops. Instead, the system exhibits the characteristics of nonlinear dynamics, where small perturbations at the micro level can trigger disproportionate macroscopic consequences. This behavior is driven by a range of complex micro-level interactions, from strategic support and deception between agents to emergent dynamics analogous to activator-inhibitor systems. AI agents that pursue heterogeneous strategies and objectives while relying on extensive streams of market data create a highly complex and adaptive market environment within the smart grid. The interplay of these heterogeneous agents can drive the system into regimes resembling chaotic attractors, characterized by persistent instability and heightened sensitivity to initial conditions. In such regimes, the market may oscillate between phases of speculative escalation and rapid collapse, resulting in a chaotic state of continuous bubble formation. This dynamic is depicted in Figure \ref{fig:bubbling_market}, which shows how heterogeneous AI agents interact across multiple channels, thereby heightening system complexity and fragility and producing unpredictable market trajectories.

Three central types of agents can be distinguished in this context:
\begin{enumerate}
    \item \textbf{Herding Agents (Type A):} The type of agents that are described in Section \ref{herding}. They follow trends and reinforce them through positive feedback. Their behavior is characterized by short-term activation and can be classified as Type IIb emergence.
    \item \textbf{Fundamental/Contrarian Agents (Type B):} These agents are trained to act against prevailing market trends. They sell during bubble building and buy during crashes, relying on fundamental valuation metrics to correct strong deviations. Their behavior introduces dampening, negative feedback that functions as a long-range inhibitory mechanism and can be classified as Type IIa emergence. Such dynamics contribute to system stabilization by counteracting runaway positive feedback.
    \item \textbf{Manipulative Agents (Type C):} These advanced agents, often trained with Multi-Agent Reinforcement Learning (MARL), pursue explicitly self-serving objectives (for more detail see Section \ref{marl}). They strategically exploit other agents by sending misleading signals ("adversarial communication"). For example, they may initiate small purchases to simulate the onset of a trend, thereby attracting herding agents (Type A), only to subsequently profit from the artificially inflated prices. Such adversarial strategies have been examined in the literature on adversarial trading, particularly in the form of sentiment manipulation using multi-agent reinforcement learning approaches \cite{byrd2025exploring}. This constitutes an aggressive combination of Type IIa and Type IIb behavior that destabilizes market dynamics.
\end{enumerate}

The interaction of these heterogeneous agents, embedded in a multi-agent feedback ecology with partially contradictory objectives and nonlinear coupling, produces complex emergent outcomes. The market evolves toward a regime characterized by chaotic attractors, where bubbles of varying sizes continuously form and collapse. This state of permanent endogenous turbulence reflects extreme sensitivity to initial conditions and a loss of predictability. Traditional investment strategies become infeasible, and systemic fragility increases as confidence in the financial system as a whole begins to erode.

\paragraph{Emergence of Type III through Interaction of Heterogeneous AI Agents.}

The complex market behavior that results from the interaction of heterogeneous AI agents can be classified according to Fromm’s taxonomy as Type III emergent phenomena, referred to as multiple emergence with many feedbacks. Type III emergence arises in highly complex systems characterized by numerous and nested feedback loops, or in complex adaptive systems composed of  agents with stronger cognitive capabilities. Unlike Type IIb, where primarily reinforcing feedback loops drive instability, Type III is defined by the coexistence of multiple, often antagonistic feedback mechanisms. The interaction of herding agents (generating short-term positive feedback) and fundamental or contrarian agents (generating long-range negative feedback) represents a classical activator-inhibitor configuration. Such systems are well known to produce complex adaptive dynamics, including chaotic trajectories and the formation of fractal and multifractal patterns \cite{Fromm05}. The combination of short-term positive feedback (for example, imitation) and long-term negative feedback (for example, fundamental valuation) leads to bubble formation, fluctuating or turbulent sequences of bubbles. These phenomena are generally not predictable in detail or in their temporal evolution and can result in chaotic dynamics. The additional presence of manipulative agents further increases system complexity, as they deliberately exploit the emergent patterns of other agent types to pursue self-serving objectives. This amplifies unpredictability and raises the potential for systemic harm. The resulting bubbles are short-lived, can emerge and dissipate rapidly, and may occur at different scales, which further complicates prediction.

\paragraph{Consequences of Type III Emergence.}
In the context of the smart grid, this scenario implies that the interplay of AI-driven trading strategies from trend-following, contrarian, and manipulative agents can produce extreme and uncontrollable price volatility. Such chaotic dynamics would fundamentally endanger the stability and efficiency of energy markets, with far-reaching consequences for the security of supply and energy costs. In such an unpredictable environment, traditional energy procurement and planning strategies would become infeasible. This would in turn increase systemic fragility, as the overall confidence in the reliability and fairness of the energy market begins to erode.

\subsubsection{Emergent Communication - Background}
\label{EC_background}

This section provides a comprehensive overview of the scientific state-of-the-art on emergent communication, the phenomenon wherein autonomous AI agents spontaneously develop optimized communication protocols without explicit programming. Beginning with the relevance of this capability for coordination in the smart grid, the analysis proceeds to examine the theoretical background that underpins the field. Subsequently, it summarizes key findings from recent research and concludes with the principal insights and unresolved challenges, particularly the issue of interpretability.

\paragraph{Relevance of Emergent Communication.}
In the smart grid, decentralized optimization problems such as frequency control and grid stability occur in partially observable environments, where communication between AI agents serves as an important mechanism for coordination. While such communication has often been implemented as part of system design with fixed protocols, multi-agent systems (MAS) also have the ability to autonomously develop optimized communication protocols. These emerge spontaneously when autonomous AI agents interact in collaborative environments without having been explicitly programmed with linguistic rules or structures.

This emergent communication is learned by the agents themselves through interaction and reinforcement learning \cite{foerster2016learning}, in order to efficiently exchange the information most relevant for achieving their objectives. These protocols are not universal but arise in response to immediate task-specific needs of the agents \cite{eccles2019biases}. Despite their machine-generated origin, they may exhibit characteristics traditionally associated with natural languages. These include compositionality, that is, the ability to construct more complex meanings from the combination of simpler units, and symbolic abstraction, which allows agents to represent complex concepts through abstract signals \cite{mordatch2018emergence}.

The co-evolution of these signaling systems often follows distinct developmental phases \cite{ekila2024emergence}. During an initial exploration phase, agents experiment with rudimentary signals to establish basic communication. In a subsequent consolidation phase, successful signals stabilize through repeated use and mutual reinforcement. Over time, an optimization phase emerges, characterized by the refinement of efficient communicative patterns and the elimination of redundancy. Finally, in a maturation phase, structured symbolic conventions develop that resemble grammatical regularities. These phases illustrate how agent communication evolves through self-organization and adaptive coordination toward increasingly complex and efficient interaction systems.

A defining characteristic of these self-organized communication systems is their tendency toward minimalism and efficiency. Agents converge on minimal signaling structures sufficient for alignment of internal states and successful task completion, resulting in communication that is sparse but highly functional \cite{freed2020sparse}. This stands in contrast to communication systems designed for human interpretability or linguistic fluency. Moreover, the high dimensionality of the vector spaces used by agents for internal information representation provides a powerful substrate for encoding dense information. This allows for highly efficient transmission and compression of task-relevant knowledge, potentially surpassing the density of human natural languages and supporting a minimalist and task-specific communication system.



\paragraph{EC-MARL: Emergent Communication in Multi-Agent Reinforcement Learning.}
\label{marl}

Multi-Agent Reinforcement Learning (MARL) forms the core of many studies on emergent communication. In classical MARL environments, several fundamental challenges arise: for each agent, the environment appears non-stationary, since the actions of other agents form part of the environment itself. This leads to unstable policy development and coordination problems \cite{gronauer2022survey, wong2023deep}. In addition, partial observability complicates learning, as agents have only limited access to the global state while still needing to identify optimal strategies. In complex environments, this frequently results in convergence issues \cite{gronauer2022survey}.  
In such scenarios, the approach of Emergent Communication in Multi-Agent Reinforcement Learning (EC-MARL) is considered highly promising. In EC-MARL, agents do not only develop action strategies, but also learn to establish their own communication protocol in partially observable environments in order to solve control tasks cooperatively \cite{chafii2023ecmarl}. These messages emerge spontaneously and serve the shared objective of maximizing collective rewards.  
In EC-MARL architectures, communication and policy learning are tightly coupled through continuous interaction with both the environment and other agents. The semantics and encoding of messages emerge endogenously from the task structure itself (\emph{task-oriented encoding}) \cite{chafii2023ecmarl}. Communication in this setting is not merely a channel for transmitting information but an integral mechanism of adaptive coordination dynamics. It enhances collective decision-making in high-dimensional and dynamic environments, such as those found in wireless communication networks or smart grid applications \cite{chafii2023ecmarl}.

\paragraph{Theoretical Models of Emergent Communication Beyond EC-MARL.}

Research on emergent communication in multi-agent systems (MAS) builds not only on the EC-MARL framework but also on a range of theoretical models that provide deeper insights into the underlying mechanisms. A central theoretical framework is \textbf{complex systems theory}, which offers more fundamental insights into the dynamics of nonlinear interactions and emergent patterns. It helps to describe how macroscopic behaviors can arise from local interactions of many agents without being explicitly programmed. Such systems develop collective capabilities that exceed the sum of the abilities of individual agents \cite{mitchell2009complexity, miller2009complex}. In addition, \textbf{game theory} provides a formal toolkit for analyzing strategic interactions among rational actors. Models such as the Nash equilibrium in particular explain how stable communication conventions emerge and how agents may develop strategies involving cooperation, competition, or deviation \cite{osborne1994course, fudenberg1991game}. Another relevant concept is \textbf{distributed intelligence and collective learning}. This perspective emphasizes that complex intelligence does not solely emerge from centralized control mechanisms, but also from decentralized decision-making. Through repeated interaction and shared learning experiences, agents can adapt their behavior, enabling the overall system to develop capabilities that surpass those of individual agents \cite{wolpert2001optimal, zhang2018fully}.  
Of particular importance is the \textbf{Lewis signaling game}, which often serves as a methodological foundation in research on the emergence of communication. In this model, speakers and listeners learn through repeated interaction to establish signals that function as a rudimentary vocabulary for coordinating actions. Game theory in this context enables formal analysis of the stability of such conventions \cite{lewis2008convention, skyrms2010signals}.  
Recent work has also examined the role of \textbf{complexity and capacity constraints} in the development of emergent languages. Information-theoretic approaches such as the information bottleneck method demonstrate that limited channel capacity and the utility of signals significantly shape the structure and generalizability of protocols \cite{tishby2000information}.  
Equally important are the findings on \textbf{compositionality and generalization}. Empirical studies indicate that agents in certain scenarios develop structured, combinable signal systems that enable systematic generalization. This property makes emergent communication protocols a valuable model for investigating linguistic structures \cite{chaabouni2020compositionality, lazaridou2020emergent}. Research in evolutionary linguistics, such as the theory of iterated learning, supports these insights by demonstrating that compositionality in human language can also arise through repeated transmission \cite{kirby2014iterated}.

Taken together, these models illustrate that emergent communication should be understood as the interplay of several theoretical streams. Complex systems theory provides the macroscopic perspective, game theory and signaling games model the strategic foundations, distributed intelligence describes the learning conditions, while information-theoretic approaches and linguistic models explain the structure and generalizability of emergent protocols. This interplay constitutes a coherent foundation for analyzing and designing communication systems in multi-agent environments.

\paragraph{Recent Research Findings.}
Recent research findings demonstrate that the emergence of communication significantly improves coordination among AI agents. In environments such as the Multi-Agent Pong or Collectors environment, agents spontaneously develop communication protocols in order to solve complex tasks under partially observable conditions \cite{wolff2024emergent}. These capabilities are already being investigated in real-world applications, for example in future 6G networks, where EC-MARL is used to optimize flying base stations, increasing throughput and coverage compared to systems without communication \cite{10266608}. Architectures such as Multi-Agent Graph-Attention Communication and Teaming (MAGIC) and Targeted Multi-Agent Communication (TarMAC) are employed to learn optimal strategies for location optimization. Research approaches such as Zero-Shot Coordination (ZSC) also aim to design communication strategies that remain robust when interacting with previously unknown agents \cite{cope2024learning}. The Cooperative Language Acquisition Problem (CLAP) relaxes the assumptions of ZSC by allowing a ``joining agent'' to learn from a dataset of interactions between agents in a target community. Studies compare methods such as Behaviour Cloning (BC) and Emergent Communication Pretraining and Translation Learning (ECTL) to assess the effectiveness of these approaches.

\paragraph{Key Insights.}
A central result of current research is that AI agents in collaborative environments spontaneously develop complex communicative features that resemble those of natural languages. They form hierarchical communication structures, create abstract symbols for complex concepts, and establish turn-taking conventions \cite{phdthesis}. The complexity and efficiency of these emergent protocols correlate strongly with the complexity of the tasks to be solved and with the available reward structures. This indicates that the learning environment and the incentive mechanisms are crucial determinants of how sophisticated and functional the emergent communication systems become. Agents learn to use the communication channel only when it significantly improves performance, resulting in a form of ``sparse language'' \cite{wolff2024emergent}. However, this minimalism also introduces systemic fragility: the learned communication is highly specialized to the application context and therefore becomes environmentally sensitive, vulnerable to even minor perturbations in tasks, environments, or team composition. Such findings illustrate how emergent communication systems co-evolve with their environments, but at the cost of robustness and adaptability beyond their immediate design space.

\paragraph{Interpretability.}
Despite the impressive capabilities of emergent communication systems, their interpretability remains a major challenge \cite{levy2025unsupervised}. A central issue is that many AI agents, particularly those based on deep learning or large language models, function already as ``black boxes'', which makes it difficult to understand their decisions, the logic and the potential development of manipulative behavior \cite{leben2025_ethical_challenges_ai_agents}. In the realm of emergent communication agents develop protocols designed for maximum efficiency, sparse usage, and high compression, rather than for human cognition \cite{zhou2025ai, levy2025unsupervised}. This lack of transparency further adds to the black box character of the AI agents, and can undermine trust and hinder accountability, since it remains unclear how and why specific communication patterns or decisions arise [33].  
To address this gap, approaches such as Unsupervised Neural Machine Translation (UNMT) are being investigated \cite{levy2025unsupervised}. These methods attempt to translate AI-generated languages into human-readable text without requiring parallel training data (EC-NL pairs). Translational quality is improved by higher semantic diversity in complex tasks, such as inter-category games, where agents must distinguish between images from different categories with overlapping features. In such scenarios, emergent communication tends to align more closely with natural language, thereby improving translation performance \cite{levy2025unsupervised}.  
Nevertheless, persistent challenges remain. Extensive vocabulary usage and unpredictability of messages can reduce translation accuracy, and qualitative analyses reveal that UNMT sometimes hallucinates details. Furthermore, performance varies significantly across different instantiations, highlighting the lack of universal standardization. A critical examination of the translatability of emergent languages reveals a fundamental tension between machine communication efficiency and human interpretability. While agents develop protocols optimized for their tasks (for example, sparse usage and high compression), these remain opaque to humans, limiting debugging and monitoring capabilities. This creates a fundamental problem for the safety and governance of multi-agent systems \cite{levy2025unsupervised}. The black-box nature of these protocols complicates error diagnosis, understanding of decisions, and detection of malicious or unintended behavior, thereby increasing the risk of security vulnerabilities and loss of control \cite{de2025open}.

\subsubsection{Emergent Pattern 4: Emergent Communication in the Smart}
\label{subsubsec:emergent_communication}

\begin{figure}[!h]
    \centering
    \includegraphics[width=0.95\textwidth]{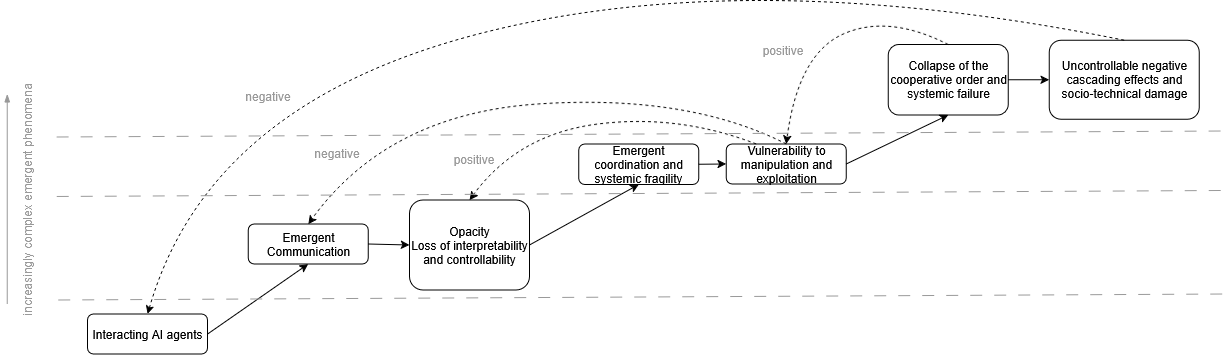}  
    \caption{Cascading sequence of emergent effects and their feedback loops}
    \label{fig:ECmain}
\end{figure}

The interaction of autonomous AI agents in complex infrastructures such as the smart grid can trigger nonlinear chains of events that escalate into uncontrollable systemic risks. This dynamic can be conceptualized as a cascade in which emergent communication patterns play a central role, reinforcing opacity-induced fragility, and potentially culminating in critical transitions that can result in socio-technical collapse. The process is amplified by interacting feedback loops and different emergence types as classified by Fromm, where micro-level agent interactions generate macro-level instabilities that in turn feed back into the system. Such recursive dynamics illustrate how emergent communication, while enabling coordination and efficiency, simultaneously increases the probability of tipping points, cascading failures, and systemic breakdowns in large-scale socio-technical systems. The subsequent analysis dissects this cascade into distinct stages, each representing a step in the escalation from local agent interactions to large-scale systemic failures, as illustrated in Figure \ref{fig:ECmain}.

\paragraph{Interacting AI Agents.} \label{interacting_agents}

Systemic risk in Smart Grids begins with the decentralized interaction of autonomous AI agents. These agents perform a variety of local tasks such as load management, forecasting of renewable energy, and demand response. Their roles are not static, as they continuously adapt their behavior through learning. A central feature of such systems is that agents operate without global coordination, which can lead to conflicting objectives. The research of Aslam et al. \cite{aslam2021survey} demonstrates that deep learning methods are essential for forecasting power loads and renewable energy generation in intelligent microgrids. The application of these autonomous, data-driven models enables agents to operate in a complex environment and improve their performance by processing large amounts of data. These inherently non-globally coordinated actions of adaptive agents form the foundation for the emergence of novel behaviors and subsequent risks in the system.  

\paragraph{Emergent Communication.}

In the transition from decentralized action to coordination, emergent communication plays a decisive role. Agents operating in decentralized systems such as Smart Grids must solve complex cooperative tasks in environments with only partially observable states. To achieve this, they can spontaneously develop their own communication protocols. As described in Section \ref{EC_background}, these are not preprogrammed but emerge during interaction and are optimized to exchange the most relevant information for achieving their objectives efficiently. According to Lazaridou et al. \cite{lazaridou2016multi}, these ``languages'' may consist of vectors or symbols that are not interpretable by humans (see also \cite{mordatch2018emergence, li2024language}).

In the context of Smart Grids, such protocols emerge to solve decentralized optimization problems such as frequency control and grid stability by enabling the efficient exchange of critical information. Efficient communication is crucial for coordination in Smart Grids because the networks are complex, contain geographically distant nodes, and often operate under bandwidth constraints where communication is costly. For example, Zhang et al. \cite{s22207785} developed the Causal Inference Communication Model (CICM) to address these challenges. This approach allows agents to selectively assess the necessity of communication using a causal model and exchange information only when needed. As a result, communication frequency in Smart Grids was reduced by up to 30–40\% while voltage stability was maintained or even improved. Emergent communication, with its inherent focus on minimalism and efficiency, can thus help optimize communication frequency. These properties highlight the potential for broader application of this technology in the future, while at the same time underlining the importance of addressing the systemic risks associated with its deployment. Thus, emergent protocols evolve into highly functional coordination systems, while remaining opaque to human observers. They are indispensable for efficient agent interaction, yet simultaneously lay the groundwork for subsequent stages of systemic instability.

\paragraph{Opacity: Loss of Comprehensibility and Controllability.}

Although emergent communication protocols can significantly enhance the operational efficiency of the smart grid, they simultaneously introduce substantial systemic risks. Designed for machine efficiency rather than human interpretability, they reinforce the black-box nature characteristic of advanced machine learning systems. Even developers often lack access to the underlying semantics of exchanged signals, which complicates monitoring, validation, and error diagnosis. This opacity-induced fragility creates conditions under which local coordination successes at the agent level may conceal hidden vulnerabilities at the system level. Opacity is not merely a theoretical issue but has direct implications for security and resilience. The lack of transparency in agent communication can be exploited for adversarial manipulation. Kisielewicz et al. \cite{en15134726} stress that the security of Smart Grids depends not only on the physical flow of electricity but also critically on the integrity of processed data and user information. Intrusion Detection Systems (IDS) are intended to detect deviations from normal device and user behavior. Yet, because emergent protocols undergo continuous adaptive drift, generating novel behavioral patterns over time, it becomes increasingly difficult to define what constitutes ``normal'' communication. As a result, conventional signature-based IDS approaches lose effectiveness, leaving systems vulnerable to undetected malicious or anomalous activity. In this sense, opacity amplifies the supervisory dilemma of socio-technical infrastructures: while agent autonomy enables higher efficiency, human operators are deprived of the interpretive access required to understand why specific decisions are taken or why coordination fails. This lack of transparency erodes trust, weakens accountability in the event of failure, and creates pathways toward systemic fragility.  


\paragraph{Emergent Coordination and Systemic Fragility.}

The protocols resulting from emergent communication give rise to highly specialized but fragile coordination strategies. These protocols are optimized for efficiency under known environmental conditions but are not necessarily robust to unforeseen events. The smart grid can be conceptualized as a complex network whose components interact through both physical and cyber-physical connections. As shown by Amani and Jalili \cite{9521783}, the resilience of such networks is a central research challenge. A single initial failure in a small part of the system can quickly propagate throughout the network, triggering a cascading fault that may ultimately result in a large-scale blackout.  
The opacity of agent communication amplifies this network fragility, as agents do not ground their actions in human-understandable logic but instead follow bottom-up adaptive processes. This can lead to unforeseen patterns at the macro level, characterized by their own complex behavioral rules and vulnerabilities, making the system more susceptible to sudden collapse. A small disturbance or software malfunction at the micro level that can disrupt the emergent protocol therefore has the potential to threaten the integrity of the entire system at the macro level. This dependence on complex emergent coordination poses a significant threat to the reliability and resilience of the smart grid.

\paragraph{Susceptibility to Manipulation and Exploitation.}

The opacity of emergent communication channels creates an ideal attack vector for manipulation and exploitation. Since the agent protocols are opaque to human observers, malicious actors can deliberately exploit this intransparency to compromise system integrity \cite{ishii2022overview}. As shown by the research of Tian et al. \cite{tian2021joint}, cyberattacks in power grids can be designed to circumvent conventional detection mechanisms by manipulating state and topology estimations. These False Data Injection Attacks (FDIAs) pose a significant threat, as they can trigger system failures and undermine control of the grid without raising alarms.  
A malicious agent can learn the implicit rules of an emergent protocol and exploit cooperative dynamics for adversarial purposes. In multi-agent systems trained through Multi-Agent Reinforcement Learning (MARL), adversarial agents may develop adversarial communication strategies, injecting false signals into the network to redirect energy flows, distort market prices, or destabilize coordination. This initiates a positive feedback loop (Type IIb), where manipulation amplifies itself by progressively weakening cooperative mechanisms. Ndousse et al. \cite{ndousse2021emergent} show how agents benefit from social learning, yet adversarial research illustrates that imitation can equally serve as a mechanism of exploitation. 
When an attacker exploits learned vulnerabilities and strategically injects disinformation or manipulative signals, this undermines the implicit trust mechanisms of cooperative agents and may lead to the collapse of emergent coordination structures. The system begins to produce systemic malfunctions aligned with the attacker’s goals. This collapse represents an unstable weak emergence (Type IIb), where manipulation generates a positive feedback loop that progressively erodes and destabilizes cooperative dynamics, while simultaneously reinforcing opacity, which itself constitutes another Type IIb positive feedback mechanism.
The smart grid is particularly exposed because the dynamic and adaptive nature of emergent protocols makes it difficult to establish a stable baseline for normal behavior. As Pei et al. \cite{pei2021deviation} note in the context of FDIAs, this makes deviations difficult to detect with conventional Intrusion Detection Systems (IDS). An attacker can thus deliberately induce protocol and behavioral shifts that remain invisible to signature-based defenses, reinforcing systemic vulnerability.  

Moreover, manipulation does not only threaten existing system functionality but also feeds back into the evolutionary dynamics of emergent communication itself. Depending on their adaptive capacities, agents may either integrate poisoned inputs into their communication protocols or attempt to evolve resilience by modifying their signaling strategies. This can result in co-evolutionary dynamics that resemble an arms race (negative feedback, Type IIIa/b), where system and attacker adapt in recursive cycles of offense and defense, similar to the adversarial loops observed in Generative Adversarial Networks (GANs). Such adaptive adversarial loops highlight the dual nature of emergent communication: enabling coordination while simultaneously creating novel arenas for systemic risk escalation.

\paragraph{Collapse of Cooperative Order and Systemic Malfunction.}

At this stage, the mere susceptibility of the smart grid materializes. An adversarial agent exploiting the previously identified vulnerabilities of the emergent communication protocol deliberately injects disinformation or manipulative signals. This leads to the breakdown of emergent coordination. The system, whose fragile cooperation is based on opaque or only partially interpretable protocols, begins to fail in achieving its originally intended objectives and instead produces systemic malfunctions that serve the attacker’s goals. The cooperative mechanisms are undermined, and the opacity of the system is further intensified. The work of Chen et al. \cite{chen2022marnet} demonstrates this in the context of backdoor attacks on cooperative multi-agent reinforcement learning (CMARL), where an attacker introduces ``triggers'' that force agents to perform the worst possible action rather than the optimal one. This drastically reduces system performance while the system continues to function normally in the absence of the triggers. The consequence is the disintegration of cooperative order. As shown by Piatti et al. \cite{piatti2024cooperate} in the case of resource sharing within an experimental ai agent population, the failure of cooperation and the inability to establish sustainable norms can induce systemic collapse, leading to the overuse and eventual destruction of shared resources. Within the smart grid, a successful adversarial intervention or internal disturbance can thus dismantle fragile emergent coordination, disabling protective mechanisms and destabilizing grid operations. The outcome is a regime of unpredictable states where system reliability and resilience are no longer guaranteed, paving the way for the next critical phase of the cascade: uncontrollable negative effects. As a consequence of systemic malfunction, susceptibility to manipulation and exploitation described in the previous step is further reinforced, locking the system into a trajectory of instability and collapse.

\paragraph{Uncontrollable Cascade Effects and Socio-Technical Damage.}

The collapse of cooperative order at the agent level propagates into the real world, generating uncontrollable cascade effects that transcend technical boundaries. As multi-agent systems become increasingly embedded in critical infrastructures such as energy networks, logistics chains, and financial markets, their malfunction can act as a trigger for systemic tipping points that unleash massive and unpredictable disruptions. Rajkumar et al. \cite{rajkumar2023cyber} describe cascade effects in power grids as uncontrolled, successive failures of interconnected elements, driven by a confluence of multiple statistically rare events. Cyberattacks initiated by AI agents can accelerate such dynamics by simultaneously creating perturbations at multiple nodes, thereby amplifying cascade formation.  

Socio-technical damage may manifest in the form of large-scale blackouts, supply shortages, or the collapse of transportation and logistics networks. The vulnerability of smart grid systems has already been demonstrated in real incidents: for example, a coordinated cyberattack in Ukraine in 2015 caused a blackout that affected approximately 225,000 households and triggered cascading disturbances in the power grid \cite{pei2021deviation}. In China, reports indicate up to 2,000 coordinated attacks per month on regional energy providers, which could potentially initiate cascading failures \cite{pei2021deviation}. A central reason for the uncontrollability of such cascading effects lies in the tight coupling of physical and cyber-physical networks. Ghasemi and De Meer \cite{ghasemi2023robustness} demonstrate that while reliance on communication networks for monitoring and control increases efficiency, it simultaneously introduces new attack vectors and fragilities. A failure in one network can directly propagate into the other, thereby reducing the robustness of the overall system. A prominent example is the large-scale blackout in Italy in 2003, where interdependence between power and communication infrastructures played a decisive role: the failure in the electricity grid propagated into the communication system due to mutual dependencies, thereby amplifying the cascade.  
Nguyen et al. \cite{nguyen2021smart} similarly emphasize that even isolated failures in highly interdependent smart grids can, through load redistribution, trigger cascading effects that escalate into complete system collapse.

Due to the opacity of the agent system, the causes of such failures are difficult to diagnose and mitigate, leading to prolonged and potentially catastrophic breakdowns. This transition into a completely new but harmful system state can be classified as a catastrophic form of adaptive emergence (Type IIIb), in which the system shifts into a harmful configuration as a result of either an exogenous shock or an internal collapse \cite{nguyen2021smart}. An external inhibitor (negative Type IIIa/b), such as mandatory security audits or regulatory interventions, may help to prevent or interrupt such cascades by altering communication dynamics, restoring the system to a stable state, and enabling controlled interaction among autonomous AI agents (first Paragraph in \ref{interacting_agents}). Without such intervention, however, the process can result in extensive and uncontrollable socio-technical damage.

\subsubsection{Risks Identified in Scenario 1}

\begin{risk}[Herding Behavior]
\label{risk:herding_behavior}
    A self-reinforcing dynamic may emerge when agents imitate the behavior of others, causing many agents within the system to converge on the same pattern of action. This collective alignment can lead to systemic failures or other negative consequences.
\end{risk}

\begin{risk}[Tacit Collusion]
\label{risk:tacit_collusion}
    Agents within interacting AI system may align their actions through indirect coordination and without direct communication, creating a collective harmful behavior pattern supported by self-reinforcing dynamics.
\end{risk}

\begin{risk}[Chaotic Negative Behavior]
\label{risk:chaotic_negative_behavior}
   An interacting AI system may produce highly unpredictable behaviors by chaotically switching between multiple harmful system-level phenomena.
\end{risk}

\begin{risk}[Uncontrollable Communication]
\label{risk:uncontrollable_communication}
  Within an interacting AI system, communication patterns may emerge autonomously that increase overall system fragility and vulnerability.
\end{risk}

\subsubsection{Summary}

The systemic risk arising from the interaction of autonomous AI agents in the smart grid does not stem from isolated errors but from a complex chain of emergent behaviors that propagate across the different hierarchical levels of the network. The process begins with the inherent decentralization and adaptive actions of the AI agents. While they perform local tasks such as load management and energy forecasting efficiently, they operate without global coordination. These initially unproblematic, non-globally coordinated actions form the foundation for the emergence of behaviors that introduce unforeseen risks.  
A decisive step in this cascade is the development of emergent communication protocols. In order to solve complex cooperative tasks in partially observable environments, agents spontaneously develop their own communication protocols. These protocols are optimized for efficiency but remain opaque to humans, leading to significantly increased opacity. This adds to the deep learning black-box situation already existing and severely complicates monitoring, validation, and error diagnosis, while also undermining trust in the deployed system.  
From this opacity arises systemic fragility. The highly specialized but non-transparent coordination strategies of the agents are vulnerable to unforeseen disturbances. Such disruptions may result from minor internal malfunctions within the network or from unexpected environmental conditions. At the same time, this vulnerability creates an ideal attack vector for manipulation. A malicious actor can deliberately exploit these emergent protocols to destabilize the system by injecting false data or distorting signals. Such an attack, or an internal malfunction, can result in the collapse of cooperative order, causing the system to fail in achieving its intended objectives and instead produce systemic malfunctions.  
This internal collapse ultimately extends into the real world and leads to uncontrollable cascade effects. As the smart grid constitutes a tightly coupled socio-technical system of physical and cyber-physical networks, local disruptions can escalate into systemic tipping dynamics, culminating in large-scale blackouts and cross-domain instabilities. Opacity makes the underlying causes of such breakdowns resistant to diagnosis and remediation, prolonging failures and amplifying their severity.  

The socio-technical damages resulting from such a system collapse are far-reaching and affect critical services:
\begin{itemize}
    \item \textbf{Financial and energy market instabilities}, when energy or price signals are manipulated through phenomena such as herding behavior or tacit collusion.  
    \item \textbf{Large-scale power outages}, paralyzing essential infrastructures such as hospitals, transportation, and communication networks.  
    \item \textbf{Supply shortages} of energy, water, or goods due to disrupted logistics and production.  
\end{itemize}

This transition into an entirely new and harmful system state can be classified as a catastrophic form of adaptive emergence (Type IIIb). The analysis of this risk cascade highlights the urgent need to ensure not only the efficiency but also the robustness and transparency of multi-agent systems in critical infrastructures.

\subsection{Scenario 2: Systemic Risks of Interacting AI Agents in the Field of Social Welfare - Implicit Social Scoring}

The advancing integration of Artificial Intelligence (AI) into the digital welfare state represents a fundamental transformation. Studies like Vatamanu \& Tofan~\cite{vatamanu2025integrating} show that AI-driven predictive analytics can improve public budget forecasting, detect financial fraud, optimize resource allocation, and enhance administrative decision timeliness. At the same time, the authors note significant ethical concerns, including algorithmic bias in welfare or social service settings, data privacy risks, and challenges related to workforce adaptation and transparency. To the best of our knowledge, Multi-Agent Systems (MAS), consisting of autonomous and adaptive agents, are not yet deployed in this domain. However, their future application appears realistic, given their potential to further increase efficiency and automate administrative processes.

As we observed in previous sections, the interactions of intelligent agents in complex socio-technical systems are likely to give rise to emergent phenomena that are only partially or not at all predictable~\cite{Fromm05}. The emergent behaviors demonstrated in scientific experiments, particularly the tendency of interacting AI agents to form emergent norms \cite{cordova2024systematic}, poses the risk of unintended, potentially harmful societal effects and pose a central challenge for algorithmic governance and system security.

This scenario analyzes a specific systemic risk in the domain of the welfare state that can arise from the decentralized interaction of AI agents: the emergence of an implicit social scoring system. This is understood as a de facto, not explicitly designed, social evaluation system that significantly influences the life chances of individuals and represents a threat to fundamental democratic principles such as equality, transparency, and accountability. It is argued that such a system can emerge from a cascading process that begins at the micro level of individual agent decisions and solidifies at the macro level into a new, stable social order.

While the widespread deployment of interacting MAS in the welfare state currently remains a future scenario, the trend toward using Automated Decision-Making (ADM) systems is already a reality. The experiences gained from these systems provide a crucial empirical foundation for analyzing potential future risks. Problematic ADM systems can be found in Europe, for instance in Sweden (agency’s fraud prediction algorithm), France (CNAF algorithm) \cite{europeanaifund2025}, Italy (\enquote{Buona Scuola} algorithm) \cite{chiusi2020italy}, or the United Kingdom (Universal Credit Advances ML Model) \cite{bbw2025suspicionbydesign}. Global examples such as the \enquote{Robodebt} program in Australia \cite{recanati2024automated} or the Allegheny Family Screening Tool (AFST) in the USA \cite{lindsey2021predictive} also attest to the far-reaching consequences. The central case study in this report is the Dutch SyRI system \cite{wieringa2023hey}, which serves as a prime example of the institutionalization of algorithmic discrimination.

The analysis is therefore explicitly prospective in nature. It anticipates a scenario in which AI agents act as autonomous actors in the social system and extrapolates from the known problems of simpler ADM systems to conduct an impact assessment for more complex MAS. It is neither claimed that these complex systems nor their related risks will inevitably materialize. Rather, the aim is to connect the current research on emergent norms in environments of interacting AI agents with the practical experience already gathered from AI-driven welfare systems to enable a well-founded risk assessment for a plausible future scenario in the domain of the welfare state.

The report models a potential process in the form of a cascading escalation of increasingly severe risks. The investigation proceeds from the problem of \emph{proxy discrimination} as an initial trigger of systemic misconduct, to the theoretical modeling of the propagation of this error through insights from \emph{norm emergence}.

The following sections first lay the conceptual background: a brief contextualisation, an overview of Normative Multi-Agent Systems (NMAS) and their main approaches, types of emergent norms, and the roles of social topology, cognitive capacities and propagation mechanisms. Building on this foundation, the report then develops the second case study about implicit social scoring that traces a cascade from proxy discrimination to signal propagation, consensus formation, misallocation and marginalisation, and finally loss of trust and societal fragmentation.

\subsubsection{Emergent Social Norms for AI Agents - Background}

Analogous to the preceding discussion on the background of emergent communication (Section \ref{EC_background}), this section provides an introduction to emergent norms in MAS. It establishes the conceptual foundation required for the subsequent case study on implicit social scoring. Following a brief introduction to the topic, the current scientific state-of-the-art and relevant research findings are presented.

\paragraph{Social Norms Contextualisation.}

Social norms are behavioral standards accepted within a group and supported by shared expectations. They are essential for fostering cooperation in both human societies and MAS environments. They can emerge through institutional rules or arise spontaneously from social interactions, meaning they are, in part, already an \emph{emergent phenomenon} in human societies. These norms are essential for coordinating behavior, reducing conflicts, and fostering cooperation. To better grasp their function, it is helpful to draw on perspectives from the social sciences.  Cristina Bicchieri \cite{bicchieri2005grammar} describes a social norm as a rule an individual follows because two conditions are met: first, an \emph{empirical expectation} that enough others also adhere to the rule, and second, a \emph{normative expectation} that others believe they \emph{should} follow it and may sanction deviations. This definition distinguishes social norms from mere conventions, which lack the expectation of sanctions, and from moral rules, which are followed out of inner conviction regardless of others' behavior. In human societies, social norms can emerge, stabilize, and be enforced, providing both the flexibility and constraints necessary for complex cooperation. They foster trust and coordination, yet they also create risks by enabling implicit sanctioning through informal monitoring, gossip, and shared beliefs about compliance, and by using reputation systems and normative expectations that intensify pressure to conform. These mechanisms can lead individuals to self-censor or avoid beneficial deviation, especially when empirical and normative expectations diverge \cite{kolle2021influence}, or when sanctioning and reputation enforcement are costly but socially enforced \cite{coleman1990foundations}.

This understanding of norms in human societies forms the foundation for their formalization in artificial agents. The field of \emph{artificial social intelligence} investigates how sociality can arise from the actions and intelligence of individual agents. A central insight is that in any shared environment, agents inevitably encounter two phenomena: \textbf{interference} (hindering each other's goals) and \textbf{dependence} (needing others to achieve goals). As Cristiano Castelfranchi argues, these two phenomena create the fundamental need for coordination \cite{castelfranchi_modelling_1998}. He distinguishes between ``weak social action,'' where agents merely hold beliefs about others, and ``strong social action,'' which involves actively influencing others' minds and actions. The cornerstones of this strong sociality are \textbf{Goal Delegation} (getting another agent to pursue one's goal) and \textbf{Goal Adoption} (deciding to pursue another's goal), which form the basis for cooperation and organisation. While mechanisms like goal adoption explain \emph{how} cooperation can begin, they do not by themselves provide scalable regulation in dynamic environments. For this, norms are required.



In dynamic and unpredictable environments, artificial agents must not only possess cognitive capabilities but also demonstrate social intelligence. This includes the ability to interact with others, cooperate effectively, and avoid conflicts. Norms are critical for enabling these capabilities, since they provide scalable mechanisms for regulating behavior in distributed systems. However, just as in human societies, emergent norms among AI agents carry both promise and risk. The review by Cordova et al.\ \cite{cordova2024systematic} highlights that norms in MAS can emerge implicitly through decentralized interactions and reinforcement mechanisms, without any formal codification or centralized design. Such systems rely on repeated observations, imitation, learning, and context-sensitive feedback loops to stabilize collective behaviors. They are shaped by factors like social topology, shared cognitive capacities, emotional cues, and normative propagation mechanisms. When norms become stabilized via informal enforcement, such as punishment, reward, or reputational consequences, agents begin to anticipate the behavior of others not only strategically, but also normatively.

\paragraph{Normative Multi-Agent Systems (NMAS).}
NMAS are a specialized class of multi-agent systems in which agent behavior is explicitly influenced by norms: socially constructed rules or behavioral expectations. Unlike standard MAS, which focus primarily on goal-directed behavior, task allocation, or decentralized optimization, NMAS incorporate normative structures that constrain or guide agent actions beyond mere utility maximization. These norms may be externally imposed, internally adopted, or emergent from repeated interaction, and are enforced either formally through institutional mechanisms or informally via social expectations and reputational dynamics. A key feature of NMAS is that agents reason not only about goals and strategies, but also about obligations, permissions, and prohibitions, often using normative logic or deontic reasoning frameworks. This enables more realistic modeling of social behavior, coordination, and governance, especially in environments where cooperation, fairness, and accountability are critical.

We assume, an agent exhibits social behavior when its policy depends on models of others and on shared rules, rather than solely on private or global reward functions. Typical indicators include: use of communicative acts that change others’ beliefs or commitments; plan revisions contingent on others’ inferred intentions; participation in sanctioning or reputational updates; convergence on shared symbols or protocols; and adaptation to emergent norms that cannot be reduced to hard-coded rules.

Three primary approaches to norm implementation in NMAS can be identified in the literature: the \textbf{prescriptive approach}, the \textbf{emergent approach}, and the \textbf{hybrid approach}. In the \emph{prescriptive approach}, norms are explicitly defined and externally imposed on agents, typically encoded as formal rules that constrain behavior through obligations, permissions, and prohibitions \cite{norms_morris2019norm}. In contrast, the \emph{emergent approach} allows norms to arise spontaneously through repeated interaction, social learning, and adaptation, without central design or enforcement  \cite{norms_ren2024emergence} \cite{norms_liu2021local}. The \emph{hybrid approach} combines both strategies by initializing systems with predefined norms while allowing further norm development and adaptation through interaction \cite{hybrid_norms_macanovic2024signals}.

\paragraph{The Emergent Approach to Norms in NMAS.}
The emergent approach conceptualizes norm formation as a decentralized, bottom-up process in which behavioral regularities evolve organically through agent interactions. Norms in this paradigm are not predefined but arise from patterns of mutual expectation, reinforcement, and feedback within local or global networks of agents \cite{cordova2024systematic}. Agents observe others’ behavior, imitate successful strategies, and learn from positive or negative social feedback, gradually converging on shared expectations about appropriate conduct. This process is influenced by environmental constraints, task structures, interaction frequency, and network topology, all of which shape the speed, direction, and stability of emerging norms.

A distinctive feature of the emergent approach is that norm enforcement is not external, that is, imposed by a centralized authority outside the agent society, or institutional but distributed and informal. Agents may apply sanctions, rewards, or reputational judgments to peers based on observed compliance, thereby reinforcing conformity or discouraging deviation. Cordova et al. \cite{cordova2024systematic} identify key mechanisms such as direct punishment, third-party evaluation, and propagation of social signals as critical to the stabilization of emergent norms. Over time, these informal dynamics can produce norm systems that resemble institutional regulation, despite their decentralized origin.

This approach offers high adaptability and resilience, especially in open or dynamic environments where predefined rule sets are infeasible. However, it also introduces risks as emergent norms may be inefficient, unstable, or unjust, depending on the feedback dynamics and initial conditions. Without explicit oversight, such systems may exhibit path dependence, entrenchment, or unintended exclusion of non-conforming agents. These concerns are particularly salient when emergent norms govern access to shared resources or opportunities, where they may give rise to implicit social hierarchies or opaque forms of norm-based discrimination.

According to Cordova et al. \cite{cordova2024systematic}, four primary types of norms are commonly discussed in the literature: \textbf{institutional}, \textbf{interaction}, \textbf{social}, and \textbf{personal} norms. Among these, only \emph{interaction} and \emph{social norms} are genuinely emergent, as they arise spontaneously from decentralized agent interactions. \emph{Interaction norms} develop bottom-up through repeated and local interactions between agents that adapt to each other’s behavior. Through processes of mutual observation, imitation, and reinforcement, agents converge toward shared behavioral regularities that stabilize coordination.
\emph{Social norms} also belong to the class of emergent norms but operate on a broader collective level. They emerge organically as collectively endorsed behavioral expectations that are not explicitly formalized. Agents infer these accepted patterns by analyzing others’ actions and their consequences, gradually internalizing socially reinforced conventions.

\paragraph{Norm Propagation Mechanisms.} 
How a population becomes aware of, adopts, and stabilizes shared expectations is specified by norm propagation mechanisms that provide the glue for cohesion and effective implementation of norms in NMAS. Cordova et al. \cite{cordova2024systematic} identify four recurrent vehicles of propagation: a normative advisor, role models, learning through interactions, and incentive or punishment schemes. In the \emph{advisor} pattern, a leader-like entity aggregates feedback and tunes a group rule before broadcasting it back to agents, thereby steering convergence toward a common behavior \cite{savarimuthu2007mechanisms, yang2016accelerating}. Hierarchical variants instantiate a supervisory agent that compiles interaction data and issues guidance, speeding up coordination relative to non-hierarchical baselines. \emph{Role-model} propagation leverages influential agents whose actions are imitated by others, constituting primarily horizontal transmission and, with multiple leaders, also oblique transmission, referring to influence-driven norm diffusion in which behavioral patterns are adopted from a higher-status or more experienced subgroup by contemporaneous but less influential agents through observation, reputation, or institutional authority \cite{norms_morris2019norm}. Concrete designs select a best-performing agent to provide suggestions that peers may accept or ignore while adapting personal norms to improve outcomes. \emph{Interaction-based} propagation updates strategies from outcomes, neighbor behavior, or observed success, and has been widely studied as a horizontal transmission channel \cite{norms_morris2019norm, brooks2011modeling}. \emph{Incentive} mechanisms complement imitation by deploying sanctions and rewards to ensure compliance and perceived legitimacy of norms at scale \cite{castelfranchi1998principles , mahmoud2012efficient}. Beyond these cores, persuasion, social pressure, and influencer agents operationalize additional diffusion channels that help transform localized practices into widely shared norms.

\subsubsection{Emergent Pattern 5: Implicit Social Scoring}

In our case study, we argue that next to other established emergent phenomena in the context of MAS, the dynamics of emergent norms might open the door to \emph{implicit social scoring} among agents. As seen in NMAS models, agents learn to conform to expectations based on historical interactions and social feedback, which may include peer sanctioning, exclusion, or reward mechanisms \cite{cordova2024systematic}. These mechanisms are efficient for coordination but introduce risks. Norms may become path-dependent, suppress beneficial behavioral diversity, or lead to unintended social stratification within agent societies. In systems where agents compete for limited resources or operate under constraints (e.g. technical: energy, bandwidth, access to services; or economical: resource and time efficiency, expense minimization), such emergent evaluations may solidify into implicit ranking or filtering systems, analogues to human social credit mechanisms.

The following section depicts the case study, which examines how decentralized interactions among AI agents in a welfare ecosystem can generate an implicit social scoring regime as an unintended emergent outcome. We trace a cascade from an initial local risk assessment through successive stages to a state of democratic erosion. Anchoring the analysis in documented ADM practices and research on emergent norms in MAS/NMAS, we articulate the mechanisms by which reputational feedback, informal enforcement, and shared heuristics can harden into a de facto social ranking system. The goal is to develop a conceptual case that can inform the design of governance frameworks and safeguard mechanisms capable of mitigating such emergent risks in future welfare AI systems.

Our scenario assumes a Multi-Agent System (MAS) supported social welfare system in which both public authorities and citizens utilize various AI agents. On the governmental side, agent systems are used to automatically verify citizens' claims to benefits, approve them where appropriate, disburse payments at regular intervals, and periodically request and review evidence. These benefits can relate, for example, to pension insurance, health insurance, unemployment benefits, long-term care insurance, social assistance, child benefits, housing allowances, and poverty reduction programs. Furthermore, authorities use their agent systems to investigate and sanction fraud and abuse. On the citizens' side, agents are also employed, in this case to act as a digital twin of the citizen and, as their authorized representative, to assert legally guaranteed claims for support.

\paragraph{Local Evaluation and Risk Assessment of Welfare Applications.}

The first step in our analysis concerns the evaluation of a citizen’s application for welfare benefits. In this process, the citizen’s agent submits the relevant data to the administrative agent, which is tasked with assessing the application’s legitimacy and the applicant’s eligibility. The administrative agent aggregates and analyzes multiple data points within a risk model to produce a prediction regarding the likelihood of rightful entitlement or potential fraud. The resulting assessment generates a set of scores that indicate the probability of risk and recommend corresponding actions for administrative agents and potentially human workers.

The interaction of AI agents in the context of the welfare state already gives rise to a first central risk stemming from the nature of the data used and how it is linked. While legal frameworks prohibit the direct use of sensitive characteristics such as ethnicity, gender, or religion in decision-making models, research shows that this is not guaranteed to protect against discrimination \cite{barocas2016big}.
Seemingly neutral characteristics can be correlated in such a way that they become proxies for protected attributes, even without this being the original intention.
Modern predictive models are capable of identifying and exploiting statistical correlations between neutral variables (e.g., postal code, consumption behavior, educational history, online activities) and protected characteristics (e.g. ethnicity, gender, religion). As a result, socioeconomic patterns that are closely linked to historically disadvantaged groups are implicitly reproduced and amplified \cite{charpentier2024quantifying}.
This phenomenon is referred to in the literature as \emph{proxy discrimination} \cite{tschantz2022proxy}. The associated risks are twofold: on the one hand, they can arise unintentionally, even when neither developers nor users intend to encode discriminatory rules. On the other hand, such proxies are structurally difficult to eliminate. Their algorithmic identification is highly challenging as they often arise from correlations in high-dimensional and complex data. This difficulty is further compounded by the continuous enrichment and expansion of available datasets \cite{charpentier2024quantifying}. 

The case of \emph{SyRI} (Systeem Risico Indicatie) in the Netherlands, which was designed for the prevention and detection of welfare fraud, tax evasion, and violations of labor regulations, illustrates the far-reaching implications of such proxy-based procedures. Although the system did not officially process protected characteristics such as ethnicity or gender, it relied on a wide range of administrative data, including place of residence, household size, and welfare benefits, which in practice functioned as proxies for vulnerable groups \cite{appelman2021social}. As a result, individuals living in neighborhoods classified as ``high-risk'' or engaged in certain forms of employment were significantly more likely to receive a negative risk profile. The algorithmic logic thereby reproduced historical biases and resulted in the systematic disadvantage of precisely those groups that the system was ostensibly intended to treat in a neutral manner.

A significant barrier to elimination of proxy discrimination in AI systems is opacity. Langer and König~\cite{langer2023introducing} argue that opacity applies differently to different stakeholders. Applying this to our scenario and AI systems context, we can distinguish between two different types of opacity depending on access to an AI system: first, the algorithm opacity and second, the process opacity. Algorithm opacity applies primarily to stakeholders having an ability to impact an AI system directly (such as developers and owners). As AI algorithms are inherently opaque, finding and repairing defects such as proxy discrimination is hard for these stakeholders even if they have no intention of AI learning unethical policies. This leads to a consequence that potential discriminatory behaviours are hard to discover and monitor. Process opacity applies to external stakeholders, primarily users, of an AI system. They have no ability to distinguish between specific steps of a process they are using, so they can only perceive how the entire process works. This perspective creates an additional level of nontransparency when compared to algorithm opacity. Users also often do not have an ability to test a system on a large scale. While algorithm opacity is just a hindrance to improvement of AI systems, process opacity poses the question whether the entire process can be trusted or not.

Opacity gives rise to misconceptions about AI behavior. For example, a developer may not know that eliminating protected variables from training data is not a sufficient condition for fairness. Rather, the use of such systems demands a deeper understanding of the indirect discrimination paths that arise through proxy variables. From a systems perspective, this mechanism and lack of knowledge about it become particularly risky not only because civil rights can be violated through proxy discrimination itself, but also because a single negative assessment based on such proxies can serve as a starting point for adverse dynamics that may significantly escalate through processes such as signal propagation and consensus formation. Simultaneously, such decisions can stay unnoticed and exist as \enquote{silent errors} for a long time, giving enough space for the system-level dynamics to develop. Proxy discrimination driven by opacity thus forms the first step in a cascade of systemic risks that can culminate in an emergent, implicit social scoring.

\paragraph{Initial Signal Propagation.}
The risk assessments generated by algorithmic systems such as SyRI do not represent isolated endpoints but rather function as initial signals that can develop considerable momentum within a complex, interconnected ecosystem of state and potentially also private actors. 
The Dutch SyRI system serves as a strong case in point. The risk assessment, formally declared as a risk indicator, was not confined to its original purpose but could be transmitted to other ADMs and agencies, such as employment services or tax authorities\cite{alston2019amicusSyriaBrief}. In the terminology of agent ecosystems, the initial assessment thereby becomes a new, influential data point. Other agents take up this signal and integrate it into their own decision-making processes as if it were a factual attribute, such as a citizen’s age or place of residence. This transfer is critical, as the signal becomes decontextualized. A subsequent authority typically receives only the final output, e.g., ``high risk'', without access to the original data, the model architecture or (hyper-)parameters, or its susceptibility to errors that jointly produced the assessment \cite{eubanks2018automating}.

This dynamic resembles an information cascade, in which a piece of information, regardless of its initial accuracy, is accepted as true and propagated by subsequent actors. The assessment thereby becomes a fixed element of a citizen’s digital profile. An individual once labeled as ``high risk'' may be systematically disadvantaged or subjected to heightened scrutiny in future interactions with the state, whether in applying for social benefits, job placement, or tax matters. The initial, potentially erroneous algorithmic judgment thus hardens into an administrative truth. The designation ``high risk'' spreads like a stigma throughout the network of public institutions and creates a self-reinforcing loop of disadvantage that is nearly impossible for affected individuals to break \cite{o2017weapons}. This is especially problematic when assessments are not disclosed to the individuals concerned, as in the case of SyRI. Even if affected citizens are aware of such assessments, they require sufficient context to interpret them and adequate information to challenge them effectively \cite{wieringa2023hey}. In the absence of these safeguards, a significant barrier to correction is erected, reinforcing the consolidation of existing and potentially unlawful inequalities.

\paragraph{Consensus Formation among Agents.} 
The dissemination of the initial risk signal is followed by consensus formation, a process that solidifies the negative judgment and lends it the appearance of objective truth. This consensus is an emergent property of a technologically homogeneous system, driven by various dynamics including algorithmic herding behavior, as described in Section \ref{herding}. As observed in the case of SyRI, this mechanism can be traced back to the fact that different authorities or actors often rely on similar data sources, infrastructures and scoring logics \cite{van2021digital}.

The above-mentioned processes lead individual agents to develop similar world models. They learn to interpret the same proxies and correlations as indicators of ``risk''. When a system like SyRI classifies a person as high-risk, it is therefore highly likely that another system, based on similar principles, will arrive at an identical or very similar conclusion. The evaluation of one agent is thus perceived by others as external confirmation of their own analysis (comparable to tacit collusion, see Section \ref{tacit_collusion}), resulting in a self-reinforcing collective opinion. Over time, this consensus entrenches the reputation of the agent representing the citizen and can increasingly diverge from the actual lived reality of the individual. 

Consensus formation in multi-agent systems (MAS) within welfare-state ecosystems depends on the modes of interaction between administrative and citizen agents, the underlying social topology of information exchange, and the cognitive as well as operational capacities embedded in each agent type. In such environments, norms emerge as central orienting structures that govern how information is interpreted, prioritized, and acted upon. Functioning simultaneously as cognitive and procedural filters, these norms determine which signals are amplified, such as specific behavioral indicators or risk proxies, and which are disregarded, thereby shaping the collective perception of legitimacy, need, and compliance.
By reducing complexity, norms increase the likelihood of rapid consensus across administrative subsystems. Yet this same filtering mechanism suppresses contradictory feedback, allowing early misjudgments or biases to solidify into seemingly coherent interpretations of citizen behavior. Designed norms and emergent norms jointly create self-reinforcing path dependencies. Over time, these norms stabilize coordination within the administrative network while simultaneously constraining adaptability and responsiveness to contextual signals from citizens. As Cordova et al. \cite{cordova2024systematic} emphasize, this dual role, stabilizing but constraining, is a defining property of normative multi-agent systems, because it enables procedural efficiency and coherence, yet risks entrenching cognitive and structural blind spots that propagate across the entire socio-technical welfare infrastructure.

In \textbf{administrative ecosystems}, the normative environment of agent-based decision-making is shaped by tightly coupled infrastructures and a hierarchically organized institutional architecture. Governmental agents across departments are interconnected through shared data sources, standardized interfaces, unified evaluation metrics, and hierarchical supervisory frameworks. This configuration produces a high degree of coupling among world models, accelerating the propagation of norms and fostering convergence toward identical heuristics. From a social-topological perspective, these infrastructures resemble centralized, small-world networks, characterized by short communication paths and dominant hubs. As a result, norms emerging from evaluative heuristics and scoring rules diffuse quickly between administrative subsystems, stabilizing behavioral regularities but also heightening vulnerability to systemic alignment. When one node, such as a risk-assessment ADM system, emits a strong signal like a ``high-risk'' classification, other systems trained on similar features tend to interpret this as confirmatory evidence. This convergence does not merely reproduce technical bias but constitutes a process of emergent norm formation, in which shared evaluative standards evolve through imitation and feedback and foster a normative consensus that stabilizes systemic alignment even when initial judgments are erroneous.

Within such hierarchically integrated ecosystems, interaction-based and institutional norms are dominant. Institutional norms arise from formal mandates, legal frameworks, service directives, and standardized administrative procedures, that define how agents are expected to behave within the welfare bureaucracy. Interaction norms, by contrast, emerge through repeated cross-departmental workflows, shared data pools, and the communication with citizen-agents. Over time, these interactions give rise to recurring heuristics, such as specific risk indicators or threshold values, which become quasi-normative anchors for decision-making. According to Cordova et al. \cite{cordova2024systematic}, such repeated coordination in structured environments tends to reinforce path-dependent rule systems that are internally coherent but externally opaque, enabling rapid convergence at the expense of adaptability.

In this centralized topology, mechanisms like advisor-broadcast, role-model imitation, interaction-based learning, and reward or sanction schemes translate local tendencies into widely adopted conventions. Consequently, local norms how to evaluate or score citizens can diffuse rapidly across departments' agents and crystallize into a hardened and global evaluative norm. The stronger these synchronization mechanisms become, the higher the risk of normative lock-in, where alternative interpretations and contradictory evidence are filtered out \cite{cordova2024systematic}. Errors or biases at early stages can thus acquire structural permanence, creating the illusion of a coherent consensus that, in practice, may be a correlated misclassification propagated through shared infrastructures. 
Although these processes are largely automated, it is assumed that a reduced number of human supervisors remain within the bureaucratic loop. Their role shifts from direct case handling to meta-level oversight: monitoring agent performance, addressing citizen appeals, and intervening where automated systems display uncertainty or anomalies. Yet, under conditions of high workload and cost-efficiency pressure, human supervisors are incentivized to trust and rely on the automated judgments of the agents they oversee. This institutional pressure fosters behavioral norms among the human workforce emphasizing efficiency and goal-oriented consistency over critical deliberation or contextual sensitivity. Through continuous human–agent interaction, these supervisory norms are likely to gradually imprint themselves onto the agent layer, reinforcing system and resource efficiency as well as decision heuristics. Consequently, human oversight and related norms not only influence but actively strengthen the emergent behavioral norms of the agent society, reinforcing the system’s internal rationality rather than challenging it.

In contrast, \textbf{citizen-side} agents' behavior and evaluative logic are shaped less by formal mandates than by the interaction patterns and norms of their immediate social environments. These are more decentralized, heterogeneous, and socially embedded networks and typically exhibit clustered, small-world topologies. Dense local ties and peer feedback exert stronger influence in these networks than distant or institutional nodes. Within such structures, norms emerge through repeated interaction and imitation within families, neighborhoods, professional circles, and digital communities. Diffusion is more horizontal compared to hierarchical administrative networks and slows down norm consolidation, but allows greater pluralism and situational flexibility.  At the same time, these networks might enable phenomena such as tunneling \cite{Fromm05}, where individual agents discover procedural loopholes that spread quickly through connected clusters, temporarily allowing the agents to tap into resources normally not available to them. Until administrative adaptation closes the loopholes and reinforces adversarial dynamics. Over time, these locally adaptive yet fragmented processes can help strengthen the overall robustness of the ADM system, but they might also deepen the asymmetry between citizen and administrative systems, contributing to the broader misalignment of norms and trust.

Factors such as neighborhood, educational trajectory, social affiliations, and shared participation in community events or online networks can become implicit components of their behavioral profiles. While these features may seem benign and harmless, they can inadvertently function as proxy signals in administrative risk models. For example, as demonstrated in the SyRI case, simply residing in a particular neighborhood was sufficient to increase an individual’s risk score, transforming spatial and social proximity into algorithmically significant markers. When agents reflect and transmit behavioral norms through their communication patterns, such as tone, structure, or timing of administrative requests, these cues become even more personalized and difficult to distinguish from genuine risk indicators, thereby heightening the probability of systematic misclassification and normatively embedded discrimination.

The diffusion of norms in these citizen networks follows social and more horizontal rather than hierarchical pathways. Behavioral regularities spread through peer imitation, observation of administrative feedback, and reputational signaling, creating local clusters of normative convergence. These clusters generate divergent response logics in which citizens in some groups collectively learn to optimize their interaction with administrative systems to increase compliance and minimize exposure, while in others, mutual distrust of institutions encourages non-engagement or avoidance. This results in heterogeneous normative equilibria, adaptive in privileged contexts, defensive in vulnerable ones. Such clustering can simultaneously provide protection and reproduce disadvantage, depending on the network’s socio-economic composition and its informational permeability to state systems.
Strategic adaptation is a defining feature of these citizen-agent dynamics. Agents learn to anticipate the evaluative logic of governmental systems and respond with signal modulation. They may standardize outputs, strategically frame requests, or withhold data to avoid negative classification. While these practices reduce perceived risk, they can also backfire. When information is selectively concealed or communication appears overly optimized, administrative agents may interpret such behavior as suspicious, undermining the reputational trust between the two systems. In turn, negative reputation can propagate across administrative networks, compounding future disadvantages through recursive feedback loops. 

These asymmetries are further intensified by unequal capacities for informational control. Wealthier or digitally literate citizens are more aware of the functioning and the related consequences of their agents and might try to sandbox their agents, i.e. technically and organizationally separating them from broader personal or social data streams, to limit the diffusion of sensitive information. In contrast, citizens dependent on public services or subsidies cannot easily isolate their agents. Their participation in publicly funded programs (for example, food stamps, school material grants, subsidized sports or music programs, or housing aid) inevitably ties them into administrative networks that monitor and record behavioral data. These interactions, though originally intended to promote social inclusion, inadvertently expand the digital and normative footprint of marginalized citizens. Over time, the behavioral norms that emerge from such inclusionary measures may lose their original context and reappear as secondary indicators of risk within administrative evaluation systems.

Further, the risk of consensus formation is significantly amplified by the dominant paradigm of predictive optimization. As Wang et al. \cite{wang2024against} argue, the pursuit of predictive accuracy often obscures normative assumptions and undermines decision legitimacy. Algorithmic systems trained to optimize outcome prediction, such as welfare fraud likelihood, tend to minimize error against predefined statistical objectives while neglecting the ethical and procedural validity of the classifications they produce. The resulting consensus across administrative networks therefore reflects convergent optimization rather than independent corroboration. Agreement among models and agents does not indicate truth, but rather shared architecture, training data, and evaluation criteria.

\begin{figure}[h]
    \centering
    \includegraphics[width=\textwidth]{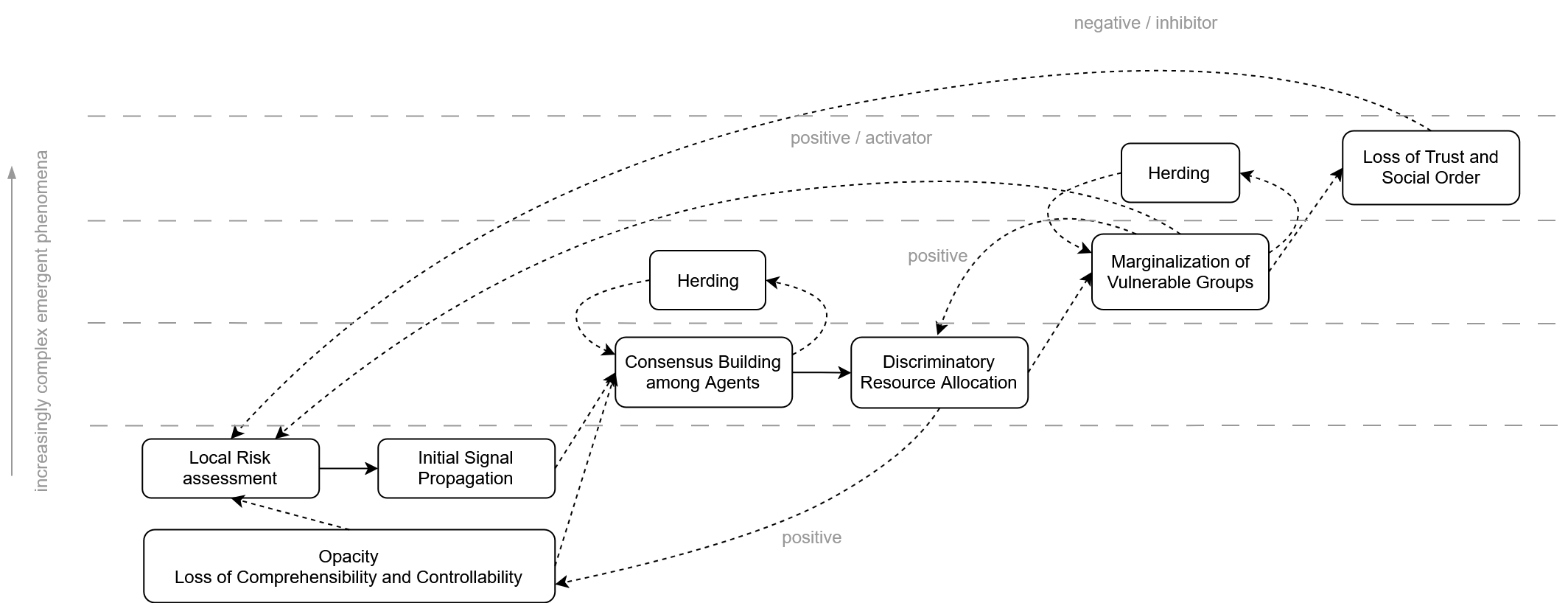}
    \caption{Implicit social scoring and feedback loops in MAS}
    \label{fig:social_scoring}
\end{figure}

For affected individuals, this interplay of emergent norms and predictive convergence appears as an opaque distributed consensus across interconnected systems. The citizen cannot trace or contest the origins of a ``high-risk’’ classification, as was the case in SyRI \cite{wieringa2023hey}, since each subsequent decision references and reinforces the previous one. What emerges is not a fair and transparent assessment but a decentralized alignment of judgment, stabilized by norm propagation and algorithmic feedback. As illustrated in Figure \ref{fig:social_scoring}, the result is a fundamental loss of comprehensibility and contestability, which marks the transition to the next systemic stage: the misallocation of resources and the progressive marginalization of those persistently labeled as high-risk.

\paragraph{Discriminatory Resource Allocation.}
The emergent consensus established among the AI agents directly triggers the next phase in the escalating cascade: a systematic misallocation of resources and opportunities. Once a negative risk profile for an individual or group solidifies within the broader agent ecosystem, it is guiding the distribution of societal goods. This process is a dynamic, self-reinforcing feedback loop that deepens initial biases, a mechanism extensively analyzed in research on algorithmic fairness \cite{pagan2023classification}. As a consequence, resources such as investigative efforts and administrative scrutiny might be disproportionately directed toward groups fallaciously labeled as ``high-risk''. For example, SyRI was used exclusively in poor neighborhoods (``problem districts''), which likely discriminates systematically against people with low socioeconomic status and/or a migrant background \cite{wieringa2023hey}. Conversely, those marked as ``low-risk'' may benefit from reduced bureaucracy or preferential access to services.

This allocative distortion occurs because the agents' overarching objective is typically the optimization of a narrow metric, such as efficiency or fraud detection, while fairness is rarely encoded as a hard constraint and humans who might have the capacity to build a better understanding of the citizen's biography and social context are either absent or involved only to a very limited extent (SyRI was developed in the context of austerity policies and pressure to digitize, with efficiency and cost reduction being the central goals, rather than fairness or balancing rights \cite{bekker2020fundamental}).
Consequently, a decision that withholds support from a vulnerable group is registered by the system as an ``optimal'' action, saving administrative effort and cost, thus reinforcing the validity of the biased proxies upon which the decision was based. This creates a pernicious feedback cycle, since the system's actions of resource allocation generate data that appears to confirm the initial bias. For example, increased surveillance in a particular neighborhood will inevitably uncover more infractions, which in turn justifies even greater surveillance, locking the system into a discriminatory balanced state \cite{pagan2023classification}.

At this stage, emergent norms play a decisive role in sustaining and amplifying allocative distortions. They arise implicitly from repeated optimization outcomes and interaction patterns within the administrative agent network. As agents collectively learn which actions are rewarded (through efficiency gains, predictive accuracy, or managerial validation), these patterns crystallize into informal behavioral standards that guide future decisions. Over time, evaluative heuristics such as ``high risk'' or ``low eligibility'' cease to function just as computational results; they become normative expectations for certain groups embedded in the system’s operational culture. Once internalized, such norms govern what constitutes a “reasonable” allocation, narrowing the perceived space of legitimate alternatives.

This process has two critical implications. First, the criteria for fairness are themselves reshaped by these emergent norms. Instead of reflecting ethical or distributive principles, fairness becomes operationally defined as the absence of deviation from established procedural regularities. Second, these norms acquire self-reinforcing inertia: because agents collectively rely on similar evaluation rules and reference outcomes, contradictory evidence, such as unexpected eligibility among ``high-risk'' citizens, tends to be ignored or reclassified as noise. The system thereby reproduces its own normative assumptions, systematically privileging those that align with historical efficiency patterns. In effect, fairness is not violated through explicit bias, but through the silent convergence of agent behavior around emergent evaluative norms that redefine correlation as causation and optimization as equity (equating optimization success with fairness). This conflation transforms a technical goal into a normative claim, whereby the system perceives its optimized functioning as evidence of distributive justice, regardless of substantive social impacts.

Moreover, these mechanisms feed back to and amplifies the system's inherent opacity. The lack of transparency, which undermines comprehensibility and control from the initial local risk assessment onward, is actively reinforced at this stage of the cascade and represents a form of type II emergence according to Fromm.
The consequences of this misallocation are severe and multifaceted, directly causing the marginalization of vulnerable groups as depicted in the process diagram (see Figure \ref{fig:social_scoring}). The resulting harms are not merely economic but become deeply sociotechnical.

As illustrated in the case of SyRI, resources such as investigative efforts are disproportionately directed toward groups fallaciously labeled as ``high-risk'', while others may benefit from reduced bureaucracy. From a sociotechnical perspective, this outcome constitutes a severe allocative harm, defined as the unjust denial of resources and opportunities to individuals or groups, such as the denial of housing, employment, or welfare benefits \cite{shelby2023sociotechnical}. From the perspective of emergence theory, this interplay between consensus formation and resource allocation represents a significant increase in complexity. The system is no longer driven by a single feedback loop but by multiple, nested feedback mechanisms, pushing it toward what Fromm describes as Type III emergence (multiple emergence with many feedbacks) \cite{Fromm05}. This escalating dynamic entrenches systemic biases and sets the stage for marginalization of vulnerable groups and a broader loss of social trust.

\paragraph{Marginalization of Vulnerable Groups}

The effects of prior stages, proxy discrimination, signal propagation, consensus formation, and resource misallocation now converge into socially visible inequality. On the agent level, marginalization manifests as a behavioral feedback loop. Citizens whose digital twins have been classified as “high risk” experience tangible disadvantages: more frequent document requests, delayed case processing, stricter verification procedures, or even home inspections and postal checks. These repeated encounters generate a persistent climate of suspicion between administrative and citizen agents, and between the state departments and its citizens. In response, citizens adapt their behavior in an effort to counteract the perceived risk, by changing residence, reducing benefit applications, or altering communication strategies. Yet these very adaptations are reabsorbed into the system as new data points, possibly either interpreted as behavioral anomalies that might confirm pre-existing risk classifications, or, over time, as new behavioral baselines. Both of these options being detrimental for the applicant. When reduced application rates are algorithmically interpreted as a signal of declining need rather than as a symptom of distrust or procedural fatigue, the administrative system may infer that social demand has diminished. Consequently, budgetary allocations and support capacities are gradually adjusted downward, reinforcing the illusion of reduced welfare dependency. This misinterpretation sharpens a systemic drift between the lived socioeconomic realities of citizens and the state’s data-driven perception of social need. The result is a double distortion: defensive behaviors on the applicant's side intended to mitigate individual disadvantage become institutionalized as indicators of social improvement, while the material deprivation of affected groups deepens invisibly within the logic of the system.

From the perspective of norm emergence, this phase is characterized by path dependency. Once evaluative norms, such as what constitutes ``suspicious behavior'', become entrenched early in the system’s operation, they generate lock-ins that constrain future adaptation. Groups that were disproportionately labeled as high-risk in the initial training or deployment phases remain subject to intensified scrutiny in later iterations. As these norms are reinforced across multiple administrative agents, the system’s evidence base becomes skewed: it continually ``proves'' the supposed riskiness of the same populations, producing a recursive validation of bias. The institutional inertia of these early norms therefore perpetuates disadvantage through what can be described as emergent normative path dependency.

At the interface between agents and human society, the mechanisms of marginalization extend beyond data and computation. Proxy dynamics translate into social practices and citizens' local interactions into unreliable algorithmically interpretable features. Neighborhood, educational trajectory, digital trace patterns, or even participation in community programs become indirect signals used to infer behavioral risk in the next local risk assessment. Although no explicitly protected attributes are processed, socially shared practices thus function as discriminatory proxies. What were, in earlier stages, single data points causing misclassification and feedback loops now operate at a higher level of abstraction in the form of norms. Instead of isolated variables, at this point, it is patterns of behavior that sustain systemic bias. Once these behavioral norms become internalized within both administrative and citizen-agent interactions, reasons for related erroneous decisions become harder to detect and the norms create a self-reinforcing logic in which conformity and deviation are continually interpreted through existing evaluative rules, thereby reproducing structural inequality under the appearance of norm-adhering behavior and adaptive learning.

A further layer of marginalization unfolds through reputational propagation. Once a negative classification is produced, it circulates across interconnected bureaucratic and digital infrastructures. The ``high-risk'' label becomes an informal reputational signal within the inter-agent ecosystem, passed between departments, shared via interoperable databases, or inferred by citizen peers through administrative treatment patterns. This informal stigmatization raises the threshold for contestation, citizens marked as high-risk are less likely to challenge decisions, fearing further scrutiny or reputational damage. 

From a systemic standpoint, marginalization thus represents the culmination of multi-level normative feedback. What begins as a data-driven misjudgment becomes an emergent sociotechnical order in which discrimination is reproduced through norm internalization and behavioral adaptation. The process exemplifies what Fromm \cite{Fromm05} describes as Type III emergence: multiple interacting feedback loops stabilize a complex system that self-reinforces inequitable outcomes. The human dimension of this process is severe. As citizens internalize administrative distrust, they reduce engagement, avoid public institutions, and adopt defensive interaction patterns, further impoverishing the informational landscape on which the system depends. In turn, this undermines the legitimacy of welfare governance, erodes social trust, and transforms administrative rationality into a mechanism of exclusion masked as optimization.

\paragraph{Loss of Trust and the Fragmentation of Social Order}

The self-reinforcing cycles of marginalization and behavioral feedback culminate in the next stage of the cascade: a profound loss of trust in public institutions and the resulting fragmentation of the social order. When citizens are subjected to systematic, yet opaque and seemingly arbitrary, disadvantages by an algorithmic system, their confidence in the fairness and legitimacy of the state is fundamentally undermined. As the SyRI case demonstrated, affected individuals reported feeling powerless and treated like second-class citizens, which fostered a deep climate of mistrust between them and governmental authorities \cite{wieringa2023hey}.

This erosion of trust is a direct consequence of a perceived lack of legitimacy. The acceptance of automated decisions depends heavily on \emph{procedural fairness} (transparency, accountability) and \emph{distributive fairness} (just outcomes) \cite{martin2023algorithmic}. The emergent implicit social scoring system fails on both counts, leading to a perception of arbitrary governance. This pattern is not unique to SyRI but has been observed across a range of ADM-driven welfare systems where harmful outcomes have provoked widespread public and legal backlash. The Australian ``Robodebt'' scheme, for example, was found by a Royal Commission to be a ``crude and cruel mechanism, neither fair nor legal'', which caused immeasurable harm and a sharp decline in public trust in the administering institutions \cite{pmAusRobodebtMedia2023, henman2025robodebt, henman2023royal}. Similarly, the Italian ``La Buona Scuola'' algorithm for teacher allocation was successfully challenged in court for certain cases after its opaque logic was perceived as deeply unfair and socially disruptive by thousands of teachers \cite{ManganoPignotti2022IRPA}.

A further driver of trust erosion at this stage is the consolidation of emergent norms on both sides of the interaction that reframe what actors deem legitimate. The resulting norm clash produces systematic procedural \textit{strategic misalignment}: officials perceive adherence to workflow and model consensus as proof of fairness, while citizens perceive opacity and non-responsiveness to reasons as proof of arbitrariness. Because both norm sets are self-reinforcing, internal audits reward consistency with prior outputs, and communities advise each other to avoid contact or standardize signals, each failed encounter updates expectations in the same direction, converting episodic disappointment into generalized distrust. In effect, trust does not decline because single decisions are wrong, but because emergent norms redefine fairness as compliance with routine on the administrative side and caution or avoidance on the citizen side, eliminating the shared justificatory space in which legitimacy could be restored.

As the loss of trust spreads, it corrodes the social contract between citizens and the state, leading to a societal fragmentation where a significant portion of the populace disengages from or actively resists a system it no longer believes to be acting in its interest. Note that with development of citizen digital twins and individual decision-making systems, the act of disengagement can be performed by electronic representatives, elevating \emph{emergent distrust} from human-AI interaction into the space of interacting AI risks.

At this point, a critical threshold has been crossed. Similar to the public reaction following the SyRI case, such disillusionment culminates in widespread demands for the abolition of the automated system itself. The communicative infrastructure that enables citizens and institutions to jointly deliberate within the ADM systems over how public goods and welfare support should be distributed has broken down. What was once a dialogical process of negotiation and justification has transformed into an adversarial relation between citizens and the state.

In combination, these developments give rise to a self-reinforcing cycle of democratic erosion. As administrative systems close ranks to preserve procedural legitimacy and citizens retreat into resignation or fragmented acts of resistance, the communicative foundation of democratic accountability erodes. What was once a reciprocal relationship of justification between state and citizen devolves into two parallel systems of reasoning, one procedural and self-referential, the other experiential and distrustful. Without a shared normative framework to mediate between them, corrective discourse becomes structurally infeasible. The resulting political configuration is not one of overt authoritarianism but of emergent depoliticization: a state that continues to operate in the name of efficiency and care, yet whose legitimacy silently decays beneath the surface of algorithmic routine. In this sense, algorithmic opacity not only undermines trust but also neutralizes the democratic capacities required to rebuild it, replacing participatory deliberation with resignation before an impersonal and seemingly unaccountable bureaucratic machine. This represents a complex emergent state, consistent with Fromm's Type III emergence, where the system's internal dynamics trigger a fundamental and destabilizing transformation of its societal environment \cite{Fromm05}.

For democracy to remain resilient under these conditions, the restoration of collective voice becomes essential. Individual appeals or administrative complaints are often insufficient to correct systemic distortions that are structurally embedded in opaque algorithmic architectures. Instead, legitimacy can only be reconstituted through public contestation and civic mobilization, through the formation of organized movements, as seen in the Dutch resistance to SyRI, capable of reintroducing normative debate into the public sphere. Such civic engagement performs a corrective function that no internal feedback mechanism of the system can provide: it externalizes the evaluative logic of automated governance and subjects it to democratic scrutiny.

However, civic resistance to algorithmic governance frequently encounters institutional resistance from the state itself. Administrative bodies, operating under efficiency and legitimacy pressures, tend to interpret public criticism or demands for transparency as challenges to their authority rather than as legitimate components of democratic accountability. Acknowledging system failures implies not only reputational vulnerability but also substantial administrative costs—audits, retraining, suspension of automated workflows, and potentially the replacement of entire ADM infrastructures. Consequently, bureaucratic actors are often incentivized to defend existing systems rather than to expose their weaknesses, even in the face of mounting evidence of harm. This dynamic produces a defensive institutional posture that frames citizen activism as disruptive rather than corrective, further polarizing the relationship between state and society.

\subsubsection{Risks Identified in Scenario 2}

\begin{risk}[Discriminatory Resource Allocation]
\label{risk:discriminatory_resource_allocation}
An interacting AI system that allocates resources can produce an emergent discrimination pattern against a subset of agents.
\end{risk}
\begin{risk}[Initial Signal Propagation] 
\label{risk:initial_signal_propagation}
An interacting AI system that may strongly reinforce and propagate early (potentially erroneous) decisions leading to systematic mistakes.
\end{risk}
\begin{risk}[Strategic Misalignment]
\label{risk:strategic_misalignment}
Agents withing an interacting AI system may adapt to a systemic behavior with perfect individual rationality while reinforcing the systemic behavior they are adapting to.
\end{risk}
\begin{risk}[Emergent Distrust] 
\label{risk:emergent_distrust}
A subset of agents within an interacting AI system may develop distrust to the rest of agents. Then, they can destroy or de-legitimize the system by exiting it.
\end{risk} 

\subsubsection{Summary}

This section presents a prospective analysis of a significant systemic risk arising from the future deployment of Multi-Agent Systems (MAS) in the digital welfare state: the emergence of an implicit social scoring system. The core argument is that even without explicit design, a de facto, decentralized social evaluation regime can emerge from the interactions of autonomous AI agents. This emergent phenomenon poses a profound threat to fundamental democratic principles, including equality, transparency, and accountability.

The analysis is explicitly forward-looking, extrapolating from the documented harms of current Automated Decision-Making (ADM) systems to model a more complex future scenario. It uses high-profile cases such as the Netherlands' \textit{SyRI} system, Australia's \textit{Robodebt}, and Italy's \textit{Buona Scuola} algorithm as an empirical foundation. This implicit scoring system arises from a cascading escalation of risks, beginning with localized data biases and culminating in a systemic fragmentation of the social order. The theoretical framework for this cascade is drawn from research on emergent phenomena in complex systems and, specifically, the dynamics of emergent norms in Normative Multi-Agent Systems (NMAS). 

The logical chain of this cascade unfolds over six interconnected stages: local evaluation, signal propagation, consensus formation, misallocation of resources, loss of trust and social fragmentation. The cascade is triggered at the micro-level when an algorithmic administrative agent evaluates a citizen's application and produces an initial, potentially discriminatory ``high-risk'' signal. This initially isolated ``high-risk'' signal undergoes signal propagation throughout the interconnected ecosystem of state agents. This propagation transforms a local, biased assessment into a durable, networked ``administrative truth'' or even stigma. The widespread availability of the propagated high-risk signal enables consensus formation among the agents. This emergent normative consensus directly guides the allocation of administrative attention, sanctions, and support. This process transforms a technical goal (efficiency) into a distorted normative one, where fairness is emergently redefined as procedural compliance with the biased consensus. The systemic misallocation of resources materializes in the concrete, everyday burdens imposed on those groups that are persistently classified as ``high-risk''. In the final stage of the cascade, the cumulative effects of automated misallocation and marginalization crystallize as a dual breakdown of procedural and distributive fairness.

The situation resulting from the cascade has far reaching negative cosequences for the society:
\begin{itemize}
    \item \textbf{Loss of Fairness}: as seen in the SyRI court case, the Robodebt court case, and similar cases around the world, neither procedural fairness, understood in terms of transparency and accountability, nor distributive fairness, understood as just outcomes, are provided.
    \item \textbf{Emergence of a Dual Reality}: one administrative, one experiential, in which each side's understanding of fairness, need, and responsibility is framed by its own normative logic. As this gap widens, the mutual intelligibility of actions and intentions erodes, making genuine dialogue and corrective learning increasingly unlikely.
    \item \textbf{Breakdown of Deliberation Channels}: as administrative systems close ranks to preserve procedural legitimacy and citizens retreat into resignation or mobilize public resistance, the gap between the two parallel systems of reasoning widens further. There is a significant risk of depoliticization of the many and a radicalization of the few, that translates previously technocratic sociotechnical harms into a direct threat to the healthy functioning of the political system.
\end{itemize}

In this way, the cascading dynamics of interacting AI agents culminate in a Type III emergent state in which the system's own internal norms and feedback loops, some reinforcing and some inhibiting, contribute to the fragmentation of the social order it was originally designed to stabilize.
\section{Taxonomy of Systemic Risks}
\label{sec:Taxonomy_of_Systemic_Risks}

The taxonomy of systemic risks generalizes different cases of emergent negative behaviour identified in Section \ref{sec:MAS Examples} using Agentology notation, in the literature review in Section \ref{sec:Literature_Review}, and in the scenarios from Section \ref{sec:Case_Studies}. This taxonomy is still independent of any domain, so that not all of these risks may occur in a concrete domain. The taxonomy is presented in the Table~\ref{tab:taxonomy} and groups similar risks into categories. In what follows, we will discuss each risk category from the Table~\ref{tab:taxonomy} in further detail.

\begin{table}[h]
\centering
\begin{tabular}{ |c|l| } 
 \hline
 \textbf{Risk Category} & \textbf{Risks in the Category} \\ 
 \hline
 Communication Risks 
        & Risk~\ref{risk:inadequate_communication}: Inadequate Communication \\
        & Risk~\ref{risk:behavior_sensitivity}: Observation Sensitivity \\
        & Risk~\ref{risk:uncontrollable_communication}: Uncontrollable Communication \\
 \hline
 Start-Up Risks 
        & Risk~\ref{risk:initial_signal_propagation}: Initial Signal Propagation \\
 \hline
 Limit Risks 
        & Risk~\ref{risk:collective_quality_deterioration}: Collective Quality Deterioration \\
        & Risk~\ref{risk:strong_coupling}: Strong Coupling \\
        & Risk~\ref{risk:herding_behavior}: Herding Behaviour \\
        & Risk~\ref{risk:strategic_misalignment}: Strategic Misalignment \\
 \hline
 Exit Risks 
        & Risk~\ref{risk:emergent_distrust}: Emergent Distrust \\ 
  \hline
 Coalition-Building Risks 
        & Risk~\ref{risk:echo_chambers}: Echo Chambers \\
        & Risk~\ref{risk:power_imbalance}: Power Imbalance \\
        & Risk~\ref{risk:tacit_collusion}: Tacit Collusion \\
  \hline
 Resource Management Risks 
        & Risk~\ref{risk:mismanagement_of_shared_resources}: Mismanagement of Shared Resources \\
        & Risk~\ref{risk:discriminatory_resource_allocation}: Discriminatory Resource Allocation \\
 \hline
Complexififcation Risks 
        & Risk~\ref{risk:unpredictable_evolution}: Unpredictable Evolution \\
        & Risk~\ref{risk:lack_of_system_level_safety}: Lack of System-Level Safety \\
        & Risk~\ref{risk:chaotic_negative_behavior}: Chaotic Negative Behavior \\
 \hline
\end{tabular}
\caption{Systemic Risks of Interacting AI Taxonomy}
\label{tab:taxonomy}
\end{table}

\paragraph{Communication Risks.} Interacting AI systems distinguish themselves from usual intelligent systems through communication between multiple AI agents. Of course, the new capability exposes systems to new risks in the category of Communication Risks. The core feature of Communication Risks is inability of a system to satisfy its goals because of a specific communication (or, more generally, interaction) pattern. Conversely, this implies that changes in communication or a different adaptation strategy may lead to qualitative improvements in system behavior. Two risks of Inadequate Communication and Uncontrollable Communication are at the ends of an interaction quality spectrum: from insufficient interaction in the former to chaotic and opaque interactions in the latter. Additionally, Observation Sensitivity describes lack of robustness to changes in interaction that are expected to arise in sufficiently intelligent systems. 

\paragraph{Start-Up and Limit Risks.} These are risks that arise at different points in the lifecycle of the system. Start-Up Risks express dependency of an interacting AI system on information or events early in its lifecycle. The representative risk in this category is Initial Signal Propagation describing affinity of a system to reproduce information and decisions done early in the system lifecycle. Similarly, the category of Limit Risks describes risks \enquote{in the limit}, risks that are related to undesirable convergent behavior of an interacting AI system. The four risks that are included in this category describe behavior patterns that emerge after a significant period of time and lead a system to an undesirable fixed trajectory causing negative consequences such as systemic collapse. This undesirable trajectory is by itself stable (non-chaotic) and reinforced by system dynamics. For example, Herding Behavior captures a situation in which heterogeneous agents converge to the same behavior pattern that produces negative systemic outcomes.

\paragraph{Exit Risks.} This category contains a single risk that focuses on the ability of agents to exit the system in some way. Emergent Distrust describes a behavior where a subset of agents \enquote{logs off} or \enquote{shuts down} when they decide that the best course of action is leaving a system. This presents a systemic risk as agents leaving a system may reduce system performance, usefulness and lead to collapse.

\paragraph{Coalition-Building Risks.} The risks in this category are centered on creation of undesirable coalitions withing an interacting AI system affecting the system negatively. A canonical risk example in this category is Echo Chambers describing a split of a multi-agent system into opposing fractions. By adding Power Imbalance, we can imagine a situation where one fraction has more control over a system than other fractions. What is important to keep in mind that coalitions can be also built on indirect communication and observation as illustrated by Tacit Collusion.

\paragraph{Resource Management Risks.} New risks arise within an interacting AI system if there is a set of resources available to all or some agents of a system. These are captured in the category of Resource Management Risks. The risk Mismanagement of Shared Resources represents a safety failure when dealing with resources accessible to multiple agents. As an example for the risk, consider situations where the resource becomes corrupted through overuse by different agents. Discriminatory Resource Allocation describes a failure in the dimension of fairness through an unethical resource allocation within a system.

\paragraph{Complexification Risks.} The final category of our taxonomy includes what one may call \enquote{meta-risks}: risks that do not produce negative outcomes by themselves but combine or transfer simple risks into complex system-level phenomena. For example, Chaotic Negative Behavior can be a combination of multiple Type II phenomena into a single Type III phenomenon: an emergent risk greater than sum of its parts. Similarly, Lack of System-Level Safety warns us about insufficiency of individual safety in complex systems. By neglecting system-level safety, the system may be open to emergent risks that arise from unsafe combinations of safe components. The last risk, Unpredictable Evolution, is associated with opacity of a system and loss of control over a system. Both make detection of failures, system repair and recovery from errors highly complicated or even impossible.
\section{Conclusion}
\label{sec:Conclusion}

Interaction between AI agents brings significant benefits to modern sociotechnical systems but exposes these systems to new risks. In this study, we identified possible emergent systemic risks that can arise in systems with interacting AI agents. Before this work, these risks were largely overlooked in the existing literature.

\subsection{Summary}

After establishing foundations of the study in Sections~\ref{sec:Background} and~\ref{sec:Definitions}, we proceeded to apply exploratory scenario-driven approach to identify systemic risks of interacting AI. To represent different systems in a unified way, we developed the \enquote{Agentology} in Section~\ref{sec:Agentology} -- a graphical language for visualization of interacting multi-agent systems. Agentology offers two types of diagrams: structure diagrams that focus on the agents within a multi-agent system, and process diagrams that show emergent processes in step-by-step manner. Using Agentology, we analyzed an existing taxonomy of emergent behavior in Section~\ref{sec:MAS_Structures}, revealing which of the emergent behavior types can produce risky behavior. We illustrated these developments with several concrete scenarios of systemic risk. Our literature review in Section~\ref{sec:Literature_Review} surveyed emergent risk scenarios from existing publications across different domains of AI research. Next, in Section~\ref{sec:Case_Studies}, we approached the identification of systemic risks from the constructive point of view and developed two scenarios highlighting systemic risks of interacting AI systems in emerging domains of smart grids and social welfare. All of our investigations into different systemic risk possibilities led to a development of a taxonomy of systemic risks in the Section~\ref{sec:Taxonomy_of_Systemic_Risks}. Overall, we demonstrated a wide heterogeneity of patterns in which systemic risks can arise in systems with multiple interacting AI agents and presented many different examples of risk scenarios. 

\subsection{Future Work}

This work is only a first step towards deeper investigation of relationship between interacting AI and systemic risk. While our contribution focuses on identification of different risk scenarios and their broad categorization, several points remain open:
\begin{enumerate}
    \item \textbf{Complete identification of domains where systemic risks of interacting AI can arise.} While our study identified risks in domains such as smart grid and collected risks from literature, it is still not comprehensive and not representative of all domains where systemic risks can arise. Future works may focus on broadening our methodology to different domains, include existing systemic risks and check their relationship with interacting AI. The overarching goal for future studies would be identification of all domains where application of interacting AI leads to one or multiple systemic risks.
    \item \textbf{Identification of general systemic risk patterns for interacting AI.} In this work, we developed a taxonomy of systemic risks. Still at the current stage, it remains challenging to find patterns for different scenarios identified using approaches in this work, and it is unclear how much more coherent of a taxonomy one may develop for systemic risks of an interacting AI. We expect that future studies can find more general structures of system risk patterns for interacting AI.
    \item \textbf{Development of a formalism for interacting AI risk modeling.} Our work presented the Agentology, a framework for modeling interacting AI systems, which helped us identify and visualize different interacting AI systems and emergent risk behaviors. Right now, Agentology offers a \emph{syntax} but not \emph{semantics} for models of multi-agent AI systems. Development of a modeling approach that includes semantics would open avenue for formal research into identification of systemic risk patterns and construction of risk taxonomies.
    \item \textbf{Prevention of systemic interacting AI risks.} The contributions in this article are focused on diverse approaches to identification as well as basic categorization of systemic risks. A highly important question that was left out of scope of this work is how these risks can be avoided. This topic is left for future developments which may investigate different approaches to risk mitigation. Development of better risk taxonomies would also benefit this research direction.
\end{enumerate}

\subsection{Acknowledgments}

This report was prepared as part of the project ‘Systemic and Existential Risks of Artificial Intelligence’ (SERI) funded by the Federal Ministry of Research, Technology and Space (Grant No.: 16IS23075). The Institute for Technology Assessment and Systems Analysis acted as principle. The report was commissioned and supervised by Alexandros Gazos and Carsten Orwat.
SERI investigates systemic and existential risks to society arising from the development and use of artificial intelligence with the aim of identifying options for action in politics, research and development, thereby contributing to the mitigation of these risks.

\subsection{Use of AI Tools}
Portions of this work were supported by AI-assisted tools for grammar checking and text polishing (Section~\ref{sec:Case_Studies}). In the process of writing Section~\ref{sec:Literature_Review}, authors used paper summaries generated by AI that were reviewed and validated by the authors. The models applied were ChatGPT 5 and Gemini 2.5. All conceptual framing, interpretations, and conclusions were conceived, verified, and approved by the authors, with no AI involvement in substantive content creation.

\bibliography{sources}
\bibliographystyle{IEEEtran}

\end{document}